%% file: content_wur_icc.tex
\documentclass[conference]{IEEEtran}
\IEEEoverridecommandlockouts
\usepackage{cite}
\usepackage{amsmath,amssymb,amsfonts}
\usepackage{algorithmic}
\usepackage{graphicx}
\usepackage{textcomp}
\usepackage{xcolor}
\def\BibTeX{{\rm B\kern-.05em{\sc i\kern-.025em b}\kern-.08em
    T\kern-.1667em\lower.7ex\hbox{E}\kern-.125emX}}
\usepackage{amssymb}
\usepackage{color}
\usepackage{graphicx}
\usepackage{amsmath}
\usepackage{cite}
\usepackage{pgfplots}
\usepackage{comment}
\usepackage{multirow}
\usepackage{mathbbol}
\usepackage{caption}
\usepackage{booktabs}
\usepackage{caption}
\usepackage{subfig}
\usepackage{tikz}
\usetikzlibrary{shapes.geometric, arrows}

\usepackage{siunitx}

\pgfplotsset{compat=newest}

\pgfplotsset{mystyle/.style={%
        width=6cm,
        xmin=0,xmax=0.5,
        xtick={0,10,...,50}}}

\newcommand{\ceil}[1]{{\left\lceil #1\right\rceil}}

\newcommand{\argmax}[1]{\underset{#1}{\operatorname{arg}\,\operatorname{max}}\;}

\newcommand{\ie}{i.e.,\,}

\newcommand{\figuresname}[1]{Figs.~}

\newcommand{\NVB}{\textcolor{blue}{$\Rightarrow$}\begin{rm} \color{blue}}   
\newcommand{\NVE}{\end{rm}\textcolor{blue}{$\Leftarrow$}}

\newif\ifcutshort
\cutshortfalse

\DeclareMathOperator{\tr}{tr}
\newcommand{\e}[1]{%
	\ifmmode\refstepcounter{equation}%
	  \eqno\mbox{\rm(\theequation)}\label{e:#1}%
	\else(\ref{e:#1})\fi}

\catcode`@=11
\newdimen\jot \jot=3pt
\def\openup{\afterassignment\@penup\dimen@=}
\def\@penup{\advance\lineskip\dimen@
  \advance\baselineskip\dimen@
  \advance\lineskiplimit\dimen@}
\def\eqalign#1{\null\,\vcenter{\openup\jot\m@th
  \ialign{\strut\hfil$\displaystyle{##}$&$\displaystyle{{}##}$\hfil
      \crcr#1\crcr}}\,}

\graphicspath{{./figure/}}

 \usepackage[acronym]{glossaries}
 \newacronym{tdma}{TDMA}{Time Division Multiple Access}
 \newacronym{ap}{AP}{Access Point}
\newacronym{ack}{ACK}{Acknowledgment}
\newacronym{dc}{DC}{Duty Cycle}
\newacronym{dr}{DR}{Data Rate}
\newacronym{dl}{DL}{downlink}
\newacronym{cca}{CCA}{Clear Channel Assessment}
\newacronym{dcw}{DCW-MAC}{Duty-Cycled Medium Access Scheme for Low-Power WUR}
\newacronym{awd}{AWD-MAC}{Asynchronous Wake-up on Demand MAC}
\newacronym{lecim}{LECIM}{Low-energy Critical Infrastructure Monitoring}
\newacronym{scm}{SCM-WUR}{Sub-Carrier Modulation Wake-up Radio Protocol}
\newacronym{opwum}{OPWUM}{Opportunistic Wake-Up MAC Protocol}
\newacronym{ed}{ED}{End Device}
\newacronym{en}{EN}{End Node}
\newacronym{rtt}{RTT}{Round Trip Time}
\newacronym{gw}{GW}{Gateway}
\newacronym{pec}{PEC}{Packet-Erasure Channel}
\newacronym{iot}{IoT}{Internet of Things}
\newacronym{ism}{ISM}{Industrial, Scientific, and Medical}
\newacronym[plural=LPWANs,firstplural=Low Power Wide Area Networks (LPWANs)]{lpwan}{LPWAN}{Low Power Wide Area Network}
\newacronym{mdp}{MDP}{Markov Decision Process}
\newacronym{mse}{MSE}{Mean Squared Error}
\newacronym{mmse}{MMSE}{Minimum MSE}
\newacronym{mcs}{MCS}{Modulation and Coding Scheme}
\newacronym{mac}{MAC}{Medium Access Control}
\newacronym{pcr}{PCR}{Primary Communication Radio}
\newacronym{phy}{PHY}{Physical}
\newacronym{pmf}{PMF}{Probability Mass Function}
\newacronym{pdf}{PDF}{Probability Density Function}
\newacronym{cdf}{CDF}{Cumulative Density Function}
\newacronym{aoi}{AoI}{Age of Information}
\newacronym{qos}{QoS}{Quality of Service}
\newacronym{rl}{RL}{Reinforcement Learning}
\newacronym{voi}{VoI}{Value of Information}
\newacronym{wsn}{WSN}{Wireless Sensor Network}
\newacronym{ul}{UL}{uplink}
\newacronym{wur}{WUR}{Wake-Up Radio}

\def \fwidth{0.8\columnwidth}
\def \fheight {0.38\columnwidth}

\def \cwidth{0.8\columnwidth}
\def \cheight {0.34\columnwidth}

\def \pwidth{0.8\columnwidth}
\def \pheight {0.2\columnwidth}

\definecolor{color1}{HTML}{FFD700}
\definecolor{color2}{HTML}{FFB14E}
\definecolor{color3}{HTML}{FA8775}
\definecolor{color4}{HTML}{EA5F94}
\definecolor{color5}{HTML}{CD34B5}
\definecolor{color6}{HTML}{9D02D7}
\definecolor{color7}{HTML}{0000FF}

\begin{document}

\title{Energy-Efficient Internet of Things Monitoring with Content-Based Wake-Up Radio}

\author{\IEEEauthorblockN{Anay Ajit Deshpande, Federico Chiariotti, Andrea Zanella}
\IEEEauthorblockA{Department of Information Engineering, University of Padova\\
Via G. Gradenigo 6/B, Padova, Italy\\
Email: \texttt{\{anayajit.deshpande,federico.chiariotti,andrea.zanella\}@unipd.it}}
\thanks{\footnotesize{This work was supported by the European Union  as part of the Italian National Recovery and Resilience Plan of NextGenerationEU, under the partnership on ``Telecommunications of the Future'' (PE0000001 - program
``RESTART'') and the ``Young Researchers'' grant REDIAL (SoE0000009).}}}

\maketitle

\begin{abstract}
The use of \gls{wur} in \gls{iot} networks can significantly improve their energy efficiency: battery-powered sensors can remain in a low-power (sleep) mode while listening for wake-up messages using their \gls{wur} and reactivate only when polled. However, polling-based \gls{wur} may still lead to wasted energy if values sensed by the polled sensors provide no new information to the receiver, or in general have a low \gls{voi}. In this paper, we design a content-based \gls{wur} that tracks the process observed by the sensors and only wakes up the sensor if its estimated update's \gls{voi} is higher than a threshold communicated through the poll. If the sensor does not reply to the polling request, the \gls{gw} can make a Bayesian update, knowing that either the sensor value substantially confirms its current estimate or the transmission failed due to the wireless channel. We analyze the trade-off between the tracking error and the battery lifetime of the sensors, showing that content-based \gls{wur} can provide fine-grained control of this trade-off and significantly increase the battery lifetime of the node with a minimal \gls{mse} increase.
\end{abstract}

\begin{IEEEkeywords}
Wake-Up Radio, Scheduling, Remote monitoring, Energy efficiency
\end{IEEEkeywords}

\section{Introduction}
\glsresetall
The explosion of the \gls{iot} has led to new developments in remote monitoring applications~\cite{wang2021evolution, zanella2023iot}, which use distributed sensors to keep track of remote environments and wide areas, as well as manufacturing plants and cities. Since the inception of the \gls{iot}, however, energy has been a major issue for system design~\cite{georgiou2017iot}: battery-powered nodes face significant constraints in terms of computational and communication capabilities, and often resort to uncoordinated random access schemes like ALOHA to avoid the signaling overhead.

However, the limits of random access schemes are well-known: unless the traffic is extremely light, these schemes suffer from packet collisions and congestion~\cite{yu2020stabilizing}, and do not allow the \gls{gw} to request new data from a specific sensor~\cite{levy1990polling}. The challenge is then to avoid the significant energy consumption incurred by nodes constantly listening for request messages, without tying the schedule to a fixed duty cycle. 

One possible solution to this problem is provided by \gls{wur} technology, standardized as part of IEEE802.11ba~\cite{deng2020ieee}: the system includes an extremely low-power radio only capable of receiving simple signals and making some basic calculations, which is kept in listening mode, while the sensor's main processor and \gls{pcr} are only turned on when needed. Typically, the \gls{wur} is used to reduce the downlink response time of a node whose \gls{pcr} is in sleep mode to save energy~\cite{zanella2023low}, achieving both a relatively low latency and a limited energy consumption. 
The standard defines the physical and \gls{mac} parameters for communication with the \gls{wur}, as well as the wake-up procedure for the \gls{pcr} when a \gls{wur} signal is received from the \gls{gw} along with the power management scheme to be implemented and associated \gls{dc} specifications and synchronization schemes. 
Crucially, it defines the channelization of wake-up frames to be sent to \gls{wur}. The standard explains the usage of ID-based wake-up messages to be sent to each node, waking up their \glspl{pcr}. 

However, the basic \gls{wur} design defined in IEEE802.11ba, hereforth referred to as ID-based \gls{wur}, does not take into account the \gls{voi} from the polled sensors. Hence, the concept of content-based \gls{wur} was proposed in~\cite{shiraishi2020content}. In content-based WUR, the polling packet carries not only the ID of the target node, but also a condition on the VoI of the data to be collected (generally, in the form of a range of interesting values). The target node then wakes the \gls{pcr} and replies to the poll on if its data satisfies the \gls{voi} requirement. Hence, in this work, we design a scheme for joint ID- and content-based \gls{wur}, defining the optimal estimate response and proposing a scheduling policy that can take into account the \gls{voi} of the update using Kalman filter estimates to increase the network lifetime while trading off some \gls{mse} performance. The proposed solution can improve the former by about 40\% in some cases, while only increasing the \gls{mse} by 10\%.

The rest of this paper is organized as follows: first, in Sec.~\ref{sec:system}, we present the basic system model for kalman filter estimation and energy consumption. We then present the censored update computation and the scheduling policy in Sec.~\ref{sec:analysis}. Results in a realistic setting are provided and discussed in Sec.~\ref{sec:results}. Sec.~\ref{sec:conc} concludes the paper and presents some possible avenues of future work on the subject.

\section{System Model}\label{sec:system}
We consider a system with $N$ distributed sensors, monitoring a physical process over a wide area. The sensors are equipped with uplink radios with wake-up functionality, and can be polled at will by the \gls{gw}. In the following, we will denote vectors using bold letters, e.g., $\mathbf{x}$, and matrices using bold capital letters, e.g., $\mathbf{A}$. Individual elements of vectors and matrices will be denoted using the same letter, with the element index as a subscript, e.g., $x_n$ or $A_{m,n}$.

\subsection{Process Model and Kalman Filter Estimation}
We model the physical process monitored by the $N$ sensors as a linear dynamic process running in discrete time. The state of the process at step $k$ is the $P\times1$ column vector $\mathbf{x}(k) = [x_1,\ldots,x_P]^T$. The dynamic system update is defined by
\begin{equation}
    \mathbf{x}(k) = \mathbf{A}\mathbf{x}(k-1) + \mathbf{v}(k),
\end{equation}
where $\mathbf{A} \in \mathbb{R}^{P\times P} $ is the update matrix for the system and $v(t)\sim \mathcal{N}(0,\mathbf{Q})$ is the Gaussian perturbation noise of the system, determined by the covariance matrix is $\mathbf{Q} \in \mathbb{R}^{P\times P}$. Each sensor $n$ then measures value $y_n(k)$, and we collect the measurements at timestep $k$ in the $N\times1$ column vector $\mathbf{y}(k)$:
\begin{equation}
    \mathbf{y}(k) = \mathbf{H}\mathbf{x}(k) + \mathbf{w}(k),
\end{equation}
where $\mathbf{H}\in\mathbb{R}^{N\times P}$ is the observation matrix and $\mathbf{w}(t) \sim \mathcal{N}(0,\mathbf{R})$ is the measurement noise vector, with covariance matrix $\mathbf{R} \in \mathbb{R}^{N\times N}$. 
As the \gls{gw} knows the process statistics, it can know or estimate $\mathbf{A}$, $\mathbf{H}$, $\mathbf{Q}$, and $\mathbf{R}$. It also has an initial estimate of the process at step 0, $\hat{\mathbf{x}}(0)$, and an initial estimation covariance matrix $\mathbf{P}(0)$, defined as:
\begin{equation}
    \mathbf{P}(0) = \mathbb{E}[(\mathbf{x}(0)-\hat{\mathbf{x}}(0))^T(\mathbf{x}(0)-\hat{\mathbf{x}}(0))].
\end{equation}
In the following, we will use the symbol $\mathbf{z}(k)$ to denote the estimation error $\mathbf{x}(k)-\hat{\mathbf{x}}(k)$. We can then use a Kalman filter, the \gls{mmse} estimator for linear processes, to obtain the \emph{a priori} estimate after each step:
\begin{equation}
\begin{cases}
    \mathbf{\hat{x}}(k|k-1) &= \mathbf{A}\hat{\mathbf{x}}(k-1);\\
    \mathbf{P}(k|k-1) &= \mathbf{A}\mathbf{P}(k-1)\mathbf{A}^T+\mathbf{Q}.
\end{cases}
\end{equation}
If sensor $n$ is polled, it can then report its measured value. Due to the wireless channel conditions, this update might be lost, in which case the \emph{a priori} estimate remains the best possible estimate of the state. Conversely, if the update is successfully received, the Kalman filter observation is simply $y_n(k)$. In the following, we will use symbol $\mathbb{1}_{n}$ to denote a one-hot row vector of length $N$, whose values are all 0 except for the $n$-th, which is equal to 1. We also define the innovation covariance matrix $\mathbf{S}(k|n)$:
\begin{equation}
\mathbf{S}(k)=\mathbf{H} \mathbf{P}(k|k-1)\mathbf{H}^T+\mathbf{R}.
\end{equation}
The Kalman gain is then:
\begin{equation}
\begin{aligned}
    \mathbf{g}(k|n) =&\frac{\mathbf{P}(k|k-1)\mathbf{H}^T\mathbb{1}_{n}^T}{S_{n,n}(k)}.
\end{aligned}
\end{equation}
The \emph{a posteriori} estimate is then:
\begin{equation}
\begin{cases}\label{eq:update_kalman}
    \hat{\mathbf{x}}(k|n)= \hat{\mathbf{x}}(k|k-1) + \mathbf{g}(k|n)\left(y_n(k)-\hat{x}_n(k|k-1)\right);\\
    \mathbf{P}(k|n) = \left(\mathbf{I}_P - \mathbf{g}(k|n)\mathbb{1}_n\mathbf{H} \right)\mathbf{P}(k|k-1),
\end{cases}
\end{equation}
where $\mathbf{I}_P$ is the $P\times P$ identity matrix.
We consider a scenario in which the process dynamics are significantly slower than the time it takes to poll a sensor, i.e., the process is much slower than polls. Hence, to avoid draining the battery of all the sensors in every single timestep, the \gls{gw} can choose to poll $M\leq N$ sensors sequentially. The updated estimate from~\eqref{eq:update_kalman} after each poll then becomes the \emph{a priori} estimate to determine the next polled sensor.

\subsection{Communication and Energy Model}

We consider a system in which one \gls{gw} communicates with $N$ wake-up capable receivers. The communication model is then a simple \gls{pec} with erasure probability $\varepsilon_n$, which includes three possible failure events:
\begin{enumerate}
\item The wake-up message might be lost due to wireless channel conditions or interference from outside the sensor network, and the sensor might not wake up and transmit its update;
\item The wake-up message might be misunderstood by another sensor, who then wakes up along with the intended node and causes a packet collision by transmitting out of turn;
\item The wireless channel conditions or interference from outside the network might not allow the \gls{gw} to correctly decode the packet transmitted by the sensors even if no other sensors transmit.
\end{enumerate}
As sensors are geographically spread out over the environment, their failure probabilities will be different. We assume that these probabilities are known to the receiver, or can be estimated beforehand.

\begin{figure}[!t]
\centering
\input{Images/id_wur_model}
    \caption{ID-based \gls{wur} sensor operation.}\vspace{-0.7cm}
    \label{fig:idbased}
\end{figure}
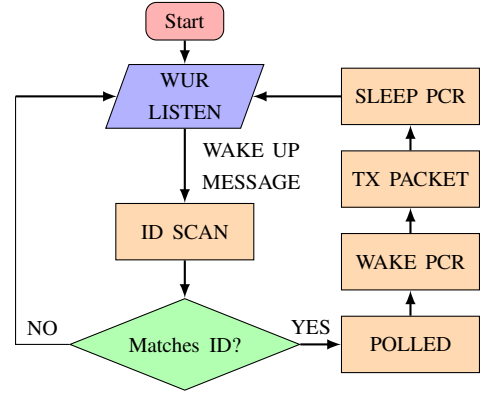

\section{Energy-Aware VoI-Based Polling}\label{sec:analysis}
Using the communication and energy model defined, we consider the three sensor operations that consume energy: the reception of a wake-up message $E_{w}$, the measurement of a new observation of the physical process monitored by the sensor $E_{s}$, and the transmission of an update $E_{t}$. \gls{wur} systems are designed to reduce the energy necessary for the wake-up radio, $E_w$, as much as possible~\cite{zaraket2021overview}. The power consumption of the \gls{wur} can be as low as \SI{2}{\micro\watt}~\cite{djidi2021can}, much lower than the power required to keep the main sensor computing unit and radio in sleep mode, which may be close to \SI{1}{\milli\watt} even for low-energy LoRa devices~\cite{nurgaliyev2020prediction}. Moreover, $E_{s}$ is usually much lower than $E_{t}$, which may require up to \SI{100}{\milli\joule} in LoRa devices, depending on the packet length and spreading factor~\cite{nurgaliyev2020prediction}.

Hence, we can consider the energy expenditure of the sensor: in a standard wake-up model, shown in Fig.~\ref{fig:idbased}, the sensor needs only a limited amount of energy $E_w$ to receive a wake-up radio message and check if its ID matches the target one, while it consumes a much larger $E_t$ to wake up the \gls{pcr} and transmit a message. 
\begin{figure}[!t]
\centering
\input{Images/content_wur_model}
    \caption{Content-based \gls{wur} sensor operation.}\vspace{-0.6cm}    
    \label{fig:contentbased}
\end{figure}
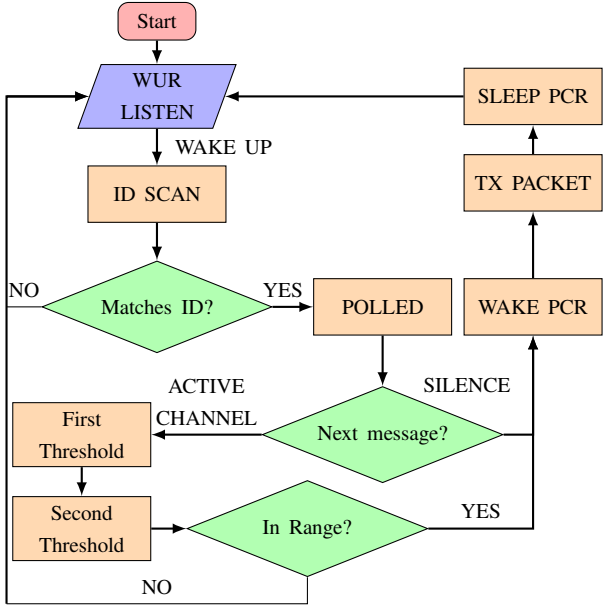
We can then envision a content-based scheme, depicted in Fig.~\ref{fig:contentbased}, which adds two more messages to the wake-up procedure. The two messages correspond to two thresholds, $a$ and $b$ which are compared to the measured value. If $b\leq a$, the sensor will wake up its \gls{pcr} and transmit any value outside the range $[b,a]$, while if $b>a$, it will transmit values only inside the interval $[a,b]$. If no further message is received after the ID message, the sensor falls back to a simple ID-based wake-up and transmits the sensed values directly. The total cost of the scheme is then $3E_w+E_s$, rather than $E_w+E_s$ as in standard \gls{wur}.

The scheme relies on sensors being able to obtain measurements using relatively low energy, and the \gls{pcr} being the most significant factor when measuring energy consumption. However, this assumption is realistic and shared by other well-known content-based \gls{wur} schemes~\cite{shiraishi2020content,kawakita2021energy,murakami2023cluster}.

\subsection{Censored Kalman Update}
As the energy consumption of the \gls{wur} is designed to be orders of magnitude lower than the \gls{pcr}'s, we can then use the content-based wake-up to improve the battery lifetime of the sensors by avoiding the transmission of updates with a lower \gls{voi}. In the following, we consider the simpler case in which $N=P$ and $\mathbf{H}=\mathbf{I}_P$, i.e., the case in which each sensor observes a component of the system state, with no influence from others, leaving the general case for future work. In particular, we consider information that confirms the \gls{gw}'s estimate to be less relevant than surprising updates that may significantly change it, \ie we set the thresholds $a=\hat{x}_n(k)-\theta$ and $b=\hat{x}_n(k)+\theta$. If the measured value $y_n(k)$ is outside the specified silent range, \ie~$y_n(k)\notin\left[\hat{x}_n(k)-\theta,\hat{x}_n(k)+\theta\right]$, the sensor transmits it, and the update occurs as described above for a normal Kalman filter. If the \gls{gw} does not receive an update, this can either be due to a channel error or to the sensor remaining silent. The probability of sensor $n$ remaining silent, an event we denote as $\xi$, is then:
\begin{equation}
\begin{aligned}
p_n(\xi)=&\Phi\left(-\theta R_{n,n}^{-\frac{1}{2}}\right)-\Phi\left(\theta R_{n,n}^{-\frac{1}{2}}\right)=1-2\Phi\left(\theta R_{n,n}^{-\frac{1}{2}}\right),
\end{aligned}
\end{equation}
where  $\Phi(x)$ is the \gls{cdf} of the standard Gaussian distribution. We can then easily compute the probability that an update is missing because the sensor was silent, and not because the packet was lost, as:
\begin{equation}
p_n(\xi|\chi)=\frac{p_n(\xi)}{p_n(\xi)+(1-p_n(\xi))\varepsilon_n},
\end{equation}
where $\chi$ indicates that there was no successful update from the sensor. The probability of having no update is then:
\begin{equation}
p_n(\chi)=\varepsilon_n+(1-\varepsilon_n)p_n(\xi).
\end{equation}
If the sensor is silent, the prior estimate $\hat{\mathbf{x}}(k|k-1)$ is simply maintained. On the other hand, we need to consider the effect of the new information on the estimate covariance. We then consider each element $P_{m,n}(k)$ in the covariance matrix $\mathbf{P}(k|k-1)$:
\begin{equation}\label{eq:pmn}
    P_{m,n}(k)=\mathbb{E}[z_n(k)z_m(k)|\mathbf{P}(k|k-1)],
\end{equation}
and then compute $P_{m,n}(k|\xi)$, i.e., each individual element of the covariance matrix in case the sensor was silent. In the following, we omit the timestep index for readability's sake. In order to compute the expected value from~\eqref{eq:pmn}, we need to apply Bayes' theorem to compute the \emph{a posteriori} \gls{pdf} of $z_n$:
\begin{equation}
p(z_n|\xi)=\frac{\phi\left(\frac{z_n}{\sqrt{P_{n,n}}}\right)\left(\Phi\left(\frac{\theta-z_n}{\sqrt{R_{n,n}}}\right)-\Phi\left(\frac{-\theta-z_n}{\sqrt{R_{n,n}}}\right)\right)}{p_n(\xi)},
\end{equation}
where $\phi(x)$ is the \gls{pdf} of the standard Gaussian distribution. By the definition of the covariance matrix, the conditional expected value of $z_m$ is simply:
\begin{equation}
\mathbb{E}[z_m|z_n]=\frac{P_{m,n}(k|k-1)z_n}{P_{n,n}(k|k-1)}.
\end{equation}
The value of $P_{m,n}(k|\xi)$ is then:
\begin{equation}
\begin{aligned}
P_{m,n}(k|\xi)=&\int_{-\infty}^{\infty}p(z_n|\xi)z_n\mathbb{E}[z_m|z_n]\ dz_n,\\
=&\frac{P_{m,n}}{P_{n,n}}\int_{-\infty}^{\infty}p(z_n|\xi)z_n^2\ dz_n.
\end{aligned}\label{eq:variance_integral}
\end{equation}
This integral does not have an analytical solution, as it involves the Gaussian \gls{cdf}, but it can be computed numerically. The calculation can then be repeated for each element of the $n$-th column of $\mathbf{P}(k)$, so as to obtain $\mathbf{P}(k|\xi)$, but the integral only needs to be solved once, as the only element that changes is $P_{m,n}(k|k-1)$. The overall update if no packet is received is:
\begin{equation}
\begin{cases}\label{eq:update_silent}
\hat{\mathbf{x}}(k|\chi)=\hat{\mathbf{x}}(k|k-1);\\
\mathbf{P}(k|\chi)=p_n(\xi|\chi)\mathbf{P}(k|\xi)+(1-p_n(\xi|\chi))\mathbf{P}(k|k-1).
\end{cases}
\end{equation}

\subsection{Poll Scheduling}
Using the system model definition, we need to define the next polling sensor that would minimize the \gls{mse} while also maximizing the network lifetime.  Hence, the scheduling strategy using the previous estimates $\{\hat{\mathbf{x}}(k-1),\mathbf{P}(k-1)\}$ can be defined as
\begin{equation}
\begin{aligned}
    a(k) = \argmax{n \in \{1,\ldots,N\} }& \tr(\mathbf{P}(k|k-1))-p_n(\chi)\tr(\mathbf{P}(k|\chi)\\
    &-(1-p_n(\chi))\tr(\mathbf{P}(k|n)).
\end{aligned}
\end{equation}
This strategy selects the sensor which offers the highest expected reduction in the overall \gls{mse}, considering the possibility of a failed or censored update. After each sensor is polled, the scheduling should be computed again with the new estimate covariance matrix, and the value of already polled sensors is set to 0.

This polling strategy is, hence, a one-step optimal heuristic~\cite{chiariotti2022scheduling}, as it greedily computes the next sensor to be polled without considering correlations. More advanced scheduling schemes that take correlations and longer-term trends into account are left for future developments.

\section{Simulation Settings and Results}\label{sec:results}
\begin{table}[!t]
\centering
\caption{Simulation Parameters}
\begin{tabular}{lcc}
\toprule
Parameter            & Symbol        & Value               \\ \midrule
Number of episodes &$L$        & 100                 \\
Timesteps per episode & $K$         & 1000                \\
Number of nodes & $N$            & 50                  \\
Number of polls per step & $M$ &\{1,2,5,10,20,50\}  \\
Value threshold & $\theta$ & $\{0.5,1,1.5,2,2.5,3\}\times\sigma$  \\ 
Transmission energy & $E_t$     & \SI{50}{\milli\joule}                \\ 
Sensing energy  & $E_{s}$        & \SI{10}{\milli\joule}                \\
Wake-up energy  & $E_{w}$ &       \SI{10}{\milli\joule}                \\
Sleep energy & $E_0$   & \SI{1}{\milli\joule}                 \\
Battery size & $E_{\max}$                 & \SI{9000}{\milli Ah}, \SI{5}{V} (\SI{162}{k\joule}) \\ \bottomrule
\end{tabular}
\label{tab:parameter}
\end{table}
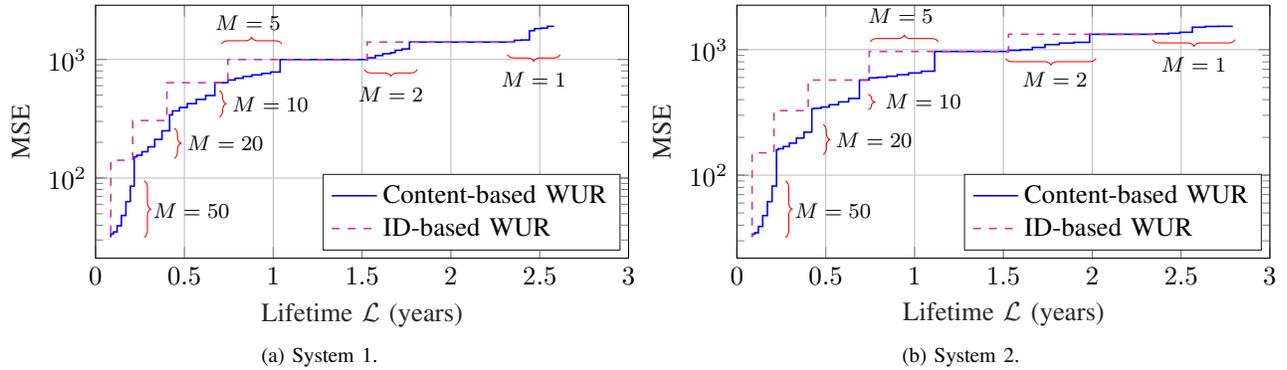
\begin{figure*}[!t]
\centering
\subfloat[%
  System 1. \label{fig:scen_1}%
]{\input{Images/Scenario_1}}
\subfloat[%
  System 2. \label{fig:scen_2}%
]{\input{Images/Scenario_2}}
    \caption{Pareto curves for the lifetime and accuracy of the schemes in the two scenarios.}\vspace{-0.7cm}
    \label{fig:results}
\end{figure*}
\begin{figure*}[!t]
\centering
\subfloat[%
  System 1. \label{fig:cdf_1}%
]{\input{Images/Scenario_1_cdf}}
\subfloat[%
  System 2. \label{fig:cdf_2}%
]{\input{Images/Scenario_2_cdf}}
    \caption{Tracking \gls{mse} \gls{cdf} with $M=10$ for the two systems.}\vspace{-0.6cm}
    \label{fig:cdf}
\end{figure*}
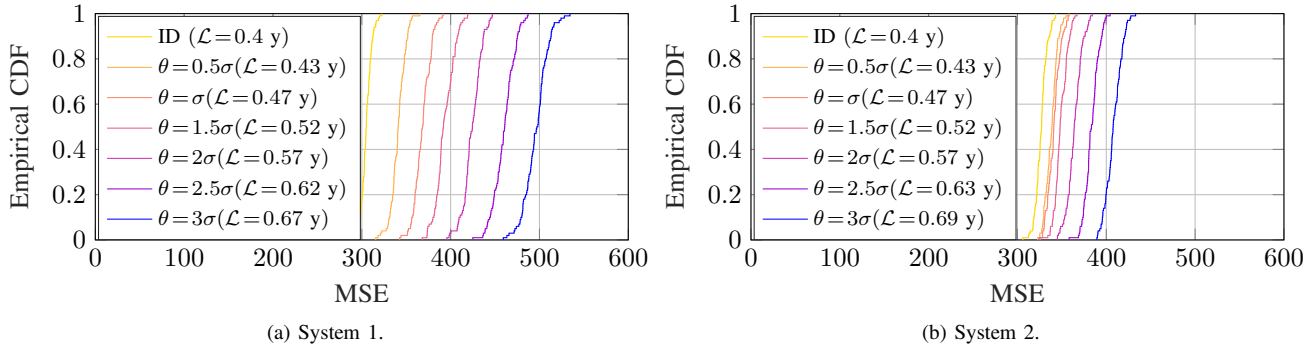
In this section, we verify the performance of the scheduling scheme by setting up a Monte Carlo simulation. We generate a stable synthetic linear process, i.e., a process whose system matrix eigenvalues are lower than 1, and apply the scheduling approach for a relatively long time.
\subsection{Simulation Settings}
We consider two different linear systems, which we allow to freely evolve over 100 episodes of 1000 timesteps each. The Monte carlo simulation is run for both classical ID-based \gls{wur} and the content-based scheme proposed in this paper, considering different values of the censoring threshold $\theta$, which is expressed as a function of the estimated \emph{a priori} uncertainty on the sensor reading, i.e., $\sqrt{P_{n,n}(k)}$ for sensor $n$. In all cases, we consider the special case in which $N=P$ and $\mathbf{H}=\mathbf{I}_P$.
We consider two systems with $N=50$ sensors for which elements of the update matrix $\mathbf{A}$ are known. In the first system, $\mathbf{A}^{(1)}$ is:
\begin{equation}
    A_{i,j}^{(1)} =
    \begin{cases}
        \frac{3}{4}, & \text{if}~i=j;\\
        -\frac{1}{8}, & \text{if}~i\neq j, \text{mod}(i-2j,4)=0;\\
        0, &\text{otherwise;}
    \end{cases}
\end{equation}
where mod($m,n$) is the integer modulo function. In the second system, $\mathbf{A}^{(2)}$ is:
\begin{equation}
    A_{i,j}^{(2)} =
    \begin{cases}
        \frac{4}{5}, & \text{if}~i=j;\\
        -\frac{1}{9}, & \text{if}~i\neq j, \text{mod}(\ceil{i-2.3j,4})=0;\\
        0, &\text{otherwise.}
    \end{cases}
\end{equation}
The other parameters remain the same for both systems. The measurement noise covariance matrix is set to $\mathbf{R}=\mathbf{I}$ and the perturbation noise covariance matrix $\mathbf{Q}$ is defined as:
\begin{equation}
    Q^{(i,j)} =
    \begin{cases}
        \frac{11+\text{mod}(i,5)}{5}, & \text{if}~i=j;\\
        1, & \text{if}~i\neq j, \text{mod}(i-j,6)=0;\\
        0,&\text{otherwise.}
    \end{cases}
\end{equation}
Note that, in both systems, the sensors with higher indices have a slightly higher variance. Additionally, the transmission error probabilities are set to $\varepsilon_n=0.02\ceil{\frac{n-1}{25}}$ and the Kalman filter is initialized at step 0 with $\hat{\mathbf{x}}(0)=\mathbf{x}(0)=0$ and $\mathbf{P}(0)=\mathbf{I}$. Overall, we particularly choose these values so as to consider two systems with distinct correlated processes: system 1 is highly interdependent, i.e., state components affect each other strongly, leading to a higher correlation, while system 2 is sparser, with a lower correlation between state components. The full simulation parameters are listed in Table~\ref{tab:parameter}.

As discussed above, the main trade-off in the system is between tracking accuracy and energy efficiency. In order to measure the former, we use the standard \gls{mse} over the state estimate and average it over all episodes:
\begin{equation}
    \text{MSE}=\frac{1}{LKN}\sum_{\ell=1}^L\sum_{k=0}^{K}(\mathbf{x}^{(\ell)}(k)-\hat{\mathbf{x}}^{(\ell)}(k))^T(\mathbf{x}^{(\ell)}(k)-\hat{\mathbf{x}}^{(\ell)}(k)),
\end{equation}
where $\mathbf{x}^{(\ell)}(k)$ denotes the state of the system in step $k$ of the $\ell$-th episode.
On the other hand, we measure energy efficiency through the sensor lifetime, i.e., the average duration of the sensor batteries:
\begin{equation}
    \mathcal{L} = \frac{1}{N}\sum_{n=1}^N\left(\frac{E_{\max}}{f_{\text{tx}} E_t+f_w(E_{s}+3E_w)+(1-f_w)E_0}\right),
\end{equation}
where $f_t(n)$ is the fraction of the total timesteps in which sensor $n$ transmitted an update and $f_w$ is the fraction of the total timesteps in which sensor $n$ was polled (including the ones in which the update was not transmitted). If we consider ID-based \gls{wur}, we have $f_t=f_w$, but the energy required to receive the wake-up signal is $E_s+E_w$ instead of $E_s+3E_w$.

Finally, we consider a timestep of 1~s for system state evaluation, in which the \gls{gw} chooses to poll a set of $M$ sensors chosen by the scheduler over a single step. We assume that $M$ is fixed over each episode, and use it as a system parameter to control the trade-off between accuracy and battery lifetime.

\subsection{Results}

The trade-off between tracking accuracy and battery lifetime can be visualized using a Pareto curve, which shows the boundary of the performance feasibility region. Any point on the curve is Pareto efficient, i.e., improving one of the performance metrics would require sacrificing the other. Fig.~\ref{fig:results} shows the Pareto curves for the two schemes in the two simulation scenarios. Each large step for ID-based \gls{wur} represents the accuracy and lifetime obtained with $M$ polled sensors, and each small step in content-based \gls{wur} represents the accuracy and lifetime achieved for a different value of $\theta$, with $M$ polled sensors. So, in this case, optimal performance would be on the lower right of the plot.
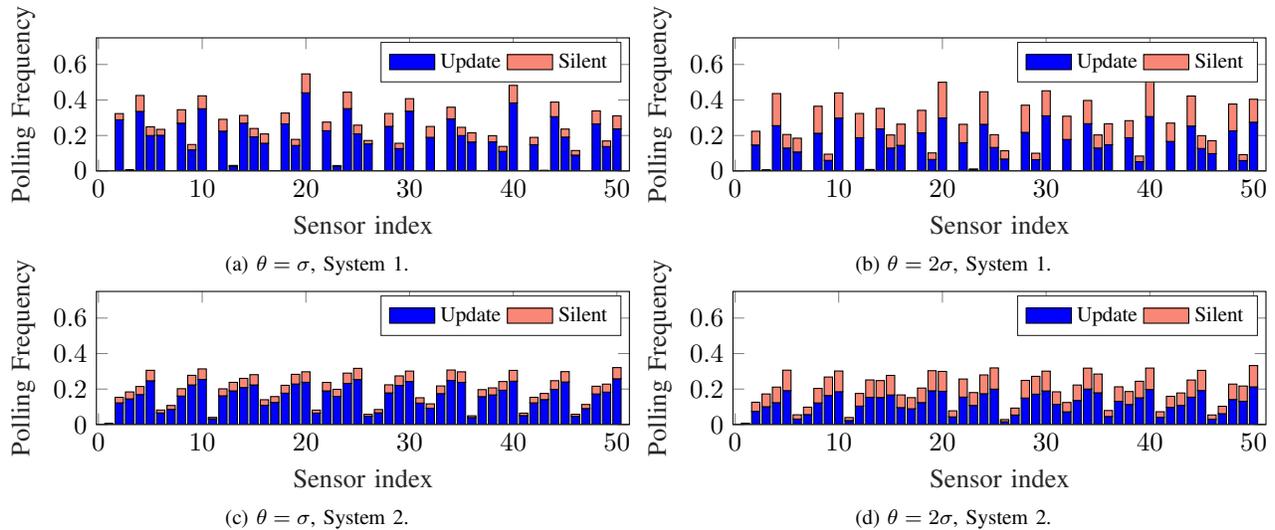
\begin{figure*}[!t]
\centering
\subfloat[%
  $\theta=\sigma$, System 1. \label{fig:selected_1_1}%
]{\input{Images/Scenario_1_selected_1}}
\subfloat[%
  $\theta=2\sigma$, System 1. \label{fig:selected_1_2}%
]{\input{Images/Scenario_1_selected_2}}\\ \vspace{-0.7cm}
\subfloat[%
  $\theta=\sigma$, System 2. \label{fig:selected_2_1}%
]{\input{Images/Scenario_2_selected_1}}
\subfloat[%
  $\theta=2\sigma$, System 2. \label{fig:selected_2_2}%
]{\input{Images/Scenario_2_selected_2}}
    \caption{Polling frequency for each sensor with $M=10$.}\vspace{-0.7cm}
    \label{fig:polling}
\end{figure*}
We can easily notice that the content-based scheme is significantly more flexible and outperforms the ID-based scheme given a fixed $M$ sensors are polled: while the two Pareto curves share some points (if we set $\theta=0$, the content-based scheme is the same as the legacy one), the content-based scheme can control the trade-off better, achieving intermediate performance points that trade some accuracy for a higher network lifetime. This is particularly evident in the second scenario, shown in Fig.~\ref{fig:scen_2}, in which increasing $\theta$ leads to a significantly smaller accuracy degradation. If we consider $M=50$, content-based \gls{wur} can increase the network lifetime by around 100\% at the cost of an \gls{mse} increase of around 50\% in both scenarios. This relative advantage diminishes for smaller values of $M$, but the absolute change in the \gls{mse} due to higher thresholds also decreases, while the battery lifetime increase is approximately the same. Additionally, the content-based scheme can reach a lifetime of over 2.5 years at the lowest accuracy setting, while the ID-based scheme would need to reduce the polling frequency below 1 poll per second to do so. 

We can analyze the accuracy-lifetime trade-off more in depth by considering the empirical \gls{cdf} of the \gls{mse} for different values of $\theta$, shown in Fig.~\ref{fig:cdf}. In this analysis, we set $M=10$, considering the intermediate part of the graph. Firstly, we can note that the network lifetime increases in a predictable fashion, as expected from the settings of the scheme: as $\theta$ is a function of the predicted sensor reading standard deviation $\sigma$, the probability of a censored update scales in a similar fashion to the Gaussian complementary \gls{cdf} function. However, there are some minor differences between the lifetimes in the two scenarios that can be attributed to the definition of $\sigma$: which is $\sqrt{P_{n,n}(k)}$ instead of $\sqrt{R_{n,n}}$, which depends on the state of the Kalman filter. 
The figures clearly show that the \gls{mse} is relatively stable across episodes and steps for all settings, as well as the trend we discussed in the two scenarios: the difference between settings is more significant in scenario 1, as the average \gls{mse} increases by 10\% for each increase in the threshold with respect to the ID-based \gls{wur} scheme's. On the other hand, scenario 2 can fare much better: setting $\theta=2\sigma$, the \gls{mse} only increases by approximately 10\%, but the network lifetime increases by more than 40\%. This is because of the structure of the two scenarios: correlation between sensors is generally lower, leading to a more uniform covariance matrix, while in the first scenario, some sensors have a much larger impact on the variance, and the consequences of a censored update from them on the \gls{mse} become more important. 

Finally, we can look at the sensor selection, shown in Fig.~\ref{fig:polling}: we can note that the polling frequency confirms that some ``central'' sensors have a relatively large impact on the overall estimation process, and therefore deplete their battery faster, while scenario 2 is much more uniform. As expected, the threshold has a limited effect on which sensors are polled, having a relatively low impact on the estimate itself, but increasing it significantly reduces the transmitted updates.
\section{Conclusion and Future Work}\label{sec:conc}
In this work, we presented a content-based \gls{wur} scheme that is able to control the trade-off between accuracy and network lifetime at a finer scale than standard ID-based \gls{wur}. To do so, the \gls{gw} transmits threshold values along with the wake-up request and the sensor only communicates when the value is outside the range, i.e., when the update's \gls{voi} is significant. The implicit communication when the sensor is silent reduces the need for explicit updates and increases the overall network lifetime, with limited \gls{mse} degradation.

The promising results shown in this paper can be extended in several direction: firstly, a dynamic optimization of the threshold values to better represent \gls{voi} may be considered. Another interesting optimization is on the schedule, which may be designed with long-term consequences in mind, considering individual sensors' batteries as well as the average lifetime.
\bibliographystyle{IEEEtran}
\bibliography{bibliography.bib}
\end{document}

%% file: Images/id_wur_model.tex
\tikzstyle{startstop} = [rectangle, rounded corners, minimum width=1cm, minimum height=0.5cm,text centered, text width=0.8cm, draw=black, fill=red!30]
\tikzstyle{io} = [trapezium, trapezium left angle=70, trapezium right angle=110, minimum width=1.5cm, minimum height=0.75cm, text centered, text width=1.25cm, draw=black, fill=blue!30]
\tikzstyle{process} = [rectangle, minimum width=1.75cm, minimum height=0.75cm, text centered, text width=1.6cm, draw=black, fill=orange!30]
\tikzstyle{decision} = [diamond, aspect=2.5, minimum width=2cm, minimum height=1cm, text centered, text width=1.75cm, draw=black, fill=green!30]
\tikzstyle{arrow} = [thick,->,>=latex, text centered, text width=1.5cm]
\begin{tikzpicture}[auto]

\node (start) [startstop] at (0,2) {\footnotesize Start};
\node (idle) [io] at (0,1) {\footnotesize WUR LISTEN};
\node (idscan) [process] at (0,-0.8) {\footnotesize ID SCAN};
\node (dec1) [decision] at (0,-2.3) {\footnotesize Matches ID?};
\node (polled) [process] at (3,-2.3) {\footnotesize POLLED};
\node (wakepcr) [process] at (3,-1.2) {\footnotesize WAKE PCR};
\node (tx) [process] at (3,-0.1) {\footnotesize TX PACKET};
\node (sleeppcr) [process] at (3,1) {\footnotesize SLEEP PCR};

\draw [arrow] (start) -- (idle);
\draw [arrow] (idle) -- node[anchor=west] {\footnotesize WAKE UP MESSAGE}(idscan);
\draw [arrow] (idscan) -- (dec1);
\draw[arrow]  (-2.25,1) -- (idle.west);
\draw (dec1.west) -| node[above,near start] {\footnotesize NO} (-2.25,1);
\draw [arrow] (dec1.east) -- node[above,near start] {\footnotesize YES} (polled.west);
\draw [arrow] (polled) -- (wakepcr);
\draw [arrow] (wakepcr) -- (tx);
\draw [arrow] (tx) -- (sleeppcr);
\draw [arrow] (sleeppcr) -- (idle);
\end{tikzpicture}

%% file: Images/content_wur_model.tex
\tikzstyle{startstop} = [rectangle, rounded corners, minimum width=1cm, minimum height=0.5cm,text centered, text width=0.8cm, draw=black, fill=red!30]
\tikzstyle{io} = [trapezium, trapezium left angle=70, trapezium right angle=110, minimum width=1.5cm, minimum height=0.75cm, text centered, text width=1.25cm, draw=black, fill=blue!30]
\tikzstyle{process} = [rectangle, minimum width=1.75cm, minimum height=0.75cm, text centered, text width=1.6cm, draw=black, fill=orange!30]
\tikzstyle{decision} = [diamond, aspect=2.5, minimum width=2cm, minimum height=1cm, text centered, text width=1.75cm, draw=black, fill=green!30]
\tikzstyle{arrow} = [thick,->,>=latex, text centered, text width=1.5cm]
\begin{tikzpicture}[auto]

\node (start) [startstop] at (0,1.5) {\footnotesize Start};
\node (idle) [io] at (0,0.5) {\footnotesize WUR LISTEN};
\node (idscan) [process] at (0,-0.8) {\footnotesize ID SCAN};
\node (dec1) [decision] at (0,-2.3) {\footnotesize Matches ID?};
\node (polled) [process] at (3,-2.3) {\footnotesize POLLED};
\node (wakepcr) [process] at (5,-2.3) {\footnotesize WAKE PCR};
\node (tx) [process] at (5,-0.65) {\footnotesize TX PACKET};
\node (sleeppcr) [process] at (5,0.5) {\footnotesize SLEEP PCR};
\node (dec2) [decision] at (3,-4) {\footnotesize Next message?};
\node (thr1) [process] at (-1,-4) {\footnotesize First Threshold};
\node (thr2) [process] at (-1,-5.25) {\footnotesize Second Threshold};
\node (dec3) [decision] at (2,-5.25) {\footnotesize In Range?};

\draw [arrow] (start) -- (idle);
\draw [arrow] (idle) -- node[anchor=west] {\footnotesize WAKE UP}(idscan);
\draw [arrow] (idscan) -- (dec1);
\draw[arrow]  (-2,0.5) -- (idle.west);
\draw (dec1.west) -| node[above,near start] {\footnotesize NO} (-2,0.5);
\draw [arrow] (dec1.east) -- node[above,near start] {\footnotesize YES} (polled.west);
\draw [arrow] (polled) -- (dec2);
\draw [arrow] (dec2) -| node[near end, left] {\footnotesize SILENCE}(wakepcr);
\draw [arrow] (dec2) -- node[midway,above] {\footnotesize ACTIVE CHANNEL}(thr1);

\draw [arrow] (wakepcr) -- (tx);
\draw [arrow] (tx) -- (sleeppcr);
\draw [arrow] (sleeppcr) -- (idle);

\draw [arrow] (dec3) -| node[near start,above] {\footnotesize YES}(wakepcr);

\draw [arrow] (thr1) -- (thr2);
\draw [arrow] (thr2) -- (dec3);
\draw (dec3) |- node[near end,above] {\footnotesize NO} (-2,-6.25);
\draw[arrow] (-2,-6.25) |- (idle);
\end{tikzpicture}

%% file: Images/Scenario_1.tex
%
%
\definecolor{mycolor1}{rgb}{0.00000,0.44700,0.74100}%
\definecolor{mycolor2}{rgb}{0.85000,0.32500,0.09800}%
\begin{tikzpicture}

\begin{semilogyaxis}[%
width=\fwidth,
height=\fheight,
at={(0.758in,0.481in)},
scale only axis,
xmin=0,
xmax=3,
xlabel style={font=\color{white!15!black}},
xlabel={Lifetime $\mathcal{L}$ (years)},
ylabel style={font=\color{white!15!black}},
ylabel={MSE},
axis background/.style={fill=white},
xmajorgrids,
ymajorgrids,
legend style={at={(0.985,0.03)}, anchor=south east, legend cell align=left, align=left, draw=white!15!black}
]
\addplot[const plot, semithick, color=color7] table[row sep=crcr] {%
2.58081862427072	1903.24324363681\\
2.54351255086908	1838.09294600732\\
2.50843347709939	1819.87030086434\\
2.47026266188264	1748.14224020704\\
2.44212374786503	1459.74447567406\\
2.39736895469855	1450.91977268207\\
2.35641573457333	1401.22574949327\\
1.76665955508463	1226.55804670361\\
1.72577033203408	1195.17269988969\\
1.68618303776412	1140.42722019449\\
1.66131165816676	1117.57563194145\\
1.61641399357267	1078.21041534074\\
1.57433774859168	1032.5868843698\\
1.52886497064579	998.160651676231\\
1.03885563010288	783.024127151953\\
0.985255706158999	762.639482570323\\
0.934405181826213	740.402463397489\\
0.879862693330065	719.015567353278\\
0.830985898626675	694.012576596312\\
0.786173216500754	670.713511407176\\
0.744490768314473	636.634883891358\\
0.671302648866413	496.194158979454\\
0.619449634639487	460.253140971688\\
0.566860984677328	424.344708545381\\
0.516188351087017	393.480538885218\\
0.472074482337738	368.31275724553\\
0.433219072492263	340.97899037906\\
0.416598244570978	250.009229967622\\
0.373778063816453	211.723144706308\\
0.332627010137003	183.189715332071\\
0.294789468419308	166.552775167709\\
0.260790953879896	155.634026139172\\
0.232101008846866	150.875606782903\\
0.217384226166198	85.357984565301\\
0.195344764097614	63.237676808296\\
0.170690649416017	48.2887743717006\\
0.145058391604583	39.4895666858472\\
0.120989804410707	35.0725539717022\\
0.100897750417863	33.8318266836473\\
0.0856164383561644	31.7182598888746\\
};
\addlegendentry{Content-based WUR}

\addplot[const plot, semithick, dashed, color=color5] table[row sep=crcr] {%
2.35641573457333	1401.22574949327\\
2.35641573457333	1401.22574949327\\
2.35641573457333	1401.22574949327\\
2.35641573457333	1401.22574949327\\
2.35641573457333	1401.22574949327\\
2.35641573457333	1401.22574949327\\
1.52886497064579	998.160651676231\\
1.52886497064579	998.160651676231\\
1.52886497064579	998.160651676231\\
1.52886497064579	998.160651676231\\
1.52886497064579	998.160651676231\\
1.52886497064579	998.160651676231\\
0.744490768314473	636.634883891358\\
0.744490768314473	636.634883891358\\
0.744490768314473	636.634883891358\\
0.744490768314473	636.634883891358\\
0.744490768314473	636.634883891358\\
0.744490768314473	636.634883891358\\
0.401327054794521	305.282301481736\\
0.401327054794521	305.282301481736\\
0.401327054794521	305.282301481736\\
0.401327054794521	305.282301481736\\
0.401327054794521	305.282301481736\\
0.401327054794521	305.282301481736\\
0.208820581356499	141.226567153165\\
0.208820581356499	141.226567153165\\
0.208820581356499	141.226567153165\\
0.208820581356499	141.226567153165\\
0.208820581356499	141.226567153165\\
0.208820581356499	141.226567153165\\
0.0856164383561644	31.7182598888746\\
0.0856164383561644	31.7182598888746\\
0.0856164383561644	31.7182598888746\\
0.0856164383561644	31.7182598888746\\
0.0856164383561644	31.7182598888746\\
0.0856164383561644	31.7182598888746\\
};
\addlegendentry{ID-based WUR}
\end{semilogyaxis}
\draw [red, decorate,
	decoration = {brace,mirror,raise=2pt}] (2.5,1.5) --  (2.5,2.25);
\node at (3.2,1.85) {\footnotesize$M=50$};
\draw [red, decorate,
	decoration = {brace,mirror,raise=2pt}] (2.9,2.55) --  (2.9,2.95);
\node at (3.65,2.75) {\footnotesize$M=20$};
\draw [red, decorate,
	decoration = {brace,mirror,raise=2pt}] (3.5,3.1) --  (3.5,3.45);
 \node at (4.25,3.27) {\footnotesize$M=10$};
\draw [red, decorate,
	decoration = {brace,raise=2pt}] (3.6,4) --  (4.4,4);
 \node at (3.95,4.35) {\footnotesize$M=5$};
 \draw [red, decorate,
	decoration = {brace,mirror, raise=2pt}] (5.5,3.8) --  (6.2,3.8);
 \node at (5.85,3.4) {\footnotesize$M=2$};
 \draw [red, decorate,
	decoration = {brace,mirror, raise=2pt}] (7.4,4) --  (8.1,4);
 \node at (7.75,3.6) {\footnotesize$M=1$};
\end{tikzpicture}%

%% file: Images/Scenario_2.tex
%
%
\definecolor{mycolor1}{rgb}{0.00000,0.44700,0.74100}%
\definecolor{mycolor2}{rgb}{0.85000,0.32500,0.09800}%
\begin{tikzpicture}[auto]

\begin{semilogyaxis}[%
width=\fwidth,
height=\fheight,
at={(0.758in,0.481in)},
scale only axis,
xmin=0,
xmax=3,
xlabel style={font=\color{white!15!black}},
xlabel={Lifetime $\mathcal{L}$ (years)},
ylabel style={font=\color{white!15!black}},
ylabel={MSE},
axis background/.style={fill=white},
xmajorgrids,
ymajorgrids,
legend style={at={(0.985,0.03)}, anchor=south east, legend cell align=left, align=left, draw=white!15!black}
]
\addplot[const plot, semithick, color=color7] table[row sep=crcr] {%
2.79185707162916	1535.23705426819\\
2.71111029438813	1531.75564428608\\
2.6355786423311	1508.89243771632\\
2.56367753879606	1376.10683771062\\
2.49331617729678	1350.30746304329\\
2.42981854659373	1337.26545474545\\
2.35641573457333	1326.22054268226\\
1.98550776505666	1143.84269362006\\
1.89658517815649	1128.7308774507\\
1.81421803369668	1097.51894088707\\
1.73413446757233	1041.40691546865\\
1.66283471358473	999.952099110219\\
1.59605766843902	987.423253535195\\
1.52886497064579	967.007801363184\\
1.11303019744776	673.012127559653\\
1.03795257365843	652.029793311875\\
0.96637955279324	631.186274016273\\
0.902210691042519	614.209296998933\\
0.843283647488087	602.486243478335\\
0.791132988219042	596.30920076241\\
0.744490768314473	572.414247699926\\
0.689685888708468	408.404855699811\\
0.628955244807722	383.640052930933\\
0.572276102351298	364.268017691635\\
0.519760440223177	348.935258524731\\
0.474009571694242	341.090957298531\\
0.434565179990987	338.316147745916\\
0.421906257663435	220.110512722953\\
0.377608569616678	197.492440777778\\
0.334981188338133	180.01539172668\\
0.295650403002436	169.258364229704\\
0.261183774731015	162.565313451529\\
0.232241170678459	159.460120489636\\
0.22297440565464	81.8656527452373\\
0.197919726032823	61.5407861205299\\
0.171580805982957	47.6332748214251\\
0.145160304435	39.0353379373858\\
0.120902374171451	34.7903023132396\\
0.100798333195924	33.8880181082127\\
0.0856164383561644	32.1254792240293\\
};
\addlegendentry{Content-based WUR}
\addplot[const plot, dashed, semithick, color=color5] table[row sep=crcr] {%
2.35641573457333	1326.22054268226\\
2.35641573457333	1326.22054268226\\
2.35641573457333	1326.22054268226\\
2.35641573457333	1326.22054268226\\
2.35641573457333	1326.22054268226\\
2.35641573457333	1326.22054268226\\
1.52886497064579	967.007801363184\\
1.52886497064579	967.007801363184\\
1.52886497064579	967.007801363184\\
1.52886497064579	967.007801363184\\
1.52886497064579	967.007801363184\\
1.52886497064579	967.007801363184\\
0.744490768314473	572.414247699926\\
0.744490768314473	572.414247699926\\
0.744490768314473	572.414247699926\\
0.744490768314473	572.414247699926\\
0.744490768314473	572.414247699926\\
0.744490768314473	572.414247699926\\
0.401327054794521	326.8074693403\\
0.401327054794521	326.8074693403\\
0.401327054794521	326.8074693403\\
0.401327054794521	326.8074693403\\
0.401327054794521	326.8074693403\\
0.401327054794521	326.8074693403\\
0.208820581356499	150.710462736192\\
0.208820581356499	150.710462736192\\
0.208820581356499	150.710462736192\\
0.208820581356499	150.710462736192\\
0.208820581356499	150.710462736192\\
0.208820581356499	150.710462736192\\
0.0856164383561644	32.1254792240293\\
0.0856164383561644	32.1254792240293\\
0.0856164383561644	32.1254792240293\\
0.0856164383561644	32.1254792240293\\
0.0856164383561644	32.1254792240293\\
0.0856164383561644	32.1254792240293\\
};
\addlegendentry{ID-based WUR}
\end{semilogyaxis}

\draw [red, decorate,
	decoration = {brace,mirror,raise=2pt}] (2.5,1.5) --  (2.5,2.25);
\node at (3.2,1.85) {\footnotesize$M=50$};
\draw [red, decorate,
	decoration = {brace,mirror,raise=2pt}] (3,2.6) --  (3,3);
\node at (3.75,2.8) {\footnotesize$M=20$};
\draw [red, decorate,
	decoration = {brace,mirror,raise=2pt}] (3.6,3.2) --  (3.6,3.4);
 \node at (4.4,3.31) {\footnotesize$M=10$};
\draw [red, decorate,
	decoration = {brace,raise=2pt}] (3.7,4.03) --  (4.6,4.03);
 \node at (4.1,4.45) {\footnotesize$M=5$};
 \draw [red, decorate,
	decoration = {brace,mirror, raise=2pt}] (5.5,3.95) --  (6.7,3.95);
 \node at (6.15,3.6) {\footnotesize$M=2$};
 \draw [red, decorate,
	decoration = {brace,mirror, raise=2pt}] (7.45,4.2) --  (8.55,4.2);
 \node at (8,3.8) {\footnotesize$M=1$};

\end{tikzpicture}%

%% file: Images/Scenario_1_cdf.tex
%
%
%
\begin{tikzpicture}

\begin{axis}[%
width=\cwidth,
height=\cheight,
at={(0.758in,0.481in)},
scale only axis,
unbounded coords=jump,
xmin=0,
xmax=600,
xlabel style={font=\color{white!15!black}},
xlabel={MSE},
ymin=0,
ymax=1,
ylabel style={font=\color{white!15!black}},
ylabel={Empirical CDF},
axis background/.style={fill=white},
xmajorgrids,
ymajorgrids,
legend style={at={(0.005,0.99)}, anchor=north west, legend cell align=left, align=left, draw=white!15!black,font={\footnotesize}}
]
\addplot [color=color1]
  table[row sep=crcr]{%
-inf	0\\
290.213522142877	0\\
290.213522142877	0.01\\
294.427000113158	0.01\\
294.427000113158	0.02\\
294.581568760792	0.02\\
294.581568760792	0.03\\
295.551941240852	0.03\\
295.551941240852	0.04\\
295.796176170401	0.04\\
295.796176170401	0.05\\
296.485559013443	0.05\\
296.485559013443	0.06\\
296.577225373133	0.06\\
296.577225373133	0.07\\
296.684593609391	0.07\\
296.684593609391	0.08\\
296.750051502558	0.08\\
296.750051502558	0.09\\
298.129631025314	0.09\\
298.129631025314	0.1\\
298.351452497448	0.1\\
298.351452497448	0.11\\
298.366944533032	0.11\\
298.366944533032	0.12\\
298.991918729584	0.12\\
298.991918729584	0.13\\
299.181263507497	0.13\\
299.181263507497	0.14\\
299.193253467699	0.14\\
299.193253467699	0.15\\
299.647529801247	0.15\\
299.647529801247	0.16\\
299.942098217845	0.16\\
299.942098217845	0.17\\
299.952306722596	0.17\\
299.952306722596	0.18\\
300.314851467826	0.18\\
300.314851467826	0.19\\
300.437526454275	0.19\\
300.437526454275	0.2\\
300.530674547751	0.2\\
300.530674547751	0.21\\
300.875807099671	0.21\\
300.875807099671	0.22\\
301.220455837096	0.22\\
301.220455837096	0.23\\
301.262284793294	0.23\\
301.262284793294	0.24\\
301.278565175976	0.24\\
301.278565175976	0.25\\
301.409825741524	0.25\\
301.409825741524	0.26\\
301.796596710787	0.26\\
301.796596710787	0.27\\
301.999685438688	0.27\\
301.999685438688	0.28\\
302.515331870149	0.28\\
302.515331870149	0.29\\
302.609185012541	0.29\\
302.609185012541	0.3\\
302.747949532707	0.3\\
302.747949532707	0.31\\
302.792973668114	0.31\\
302.792973668114	0.32\\
302.848985128363	0.32\\
302.848985128363	0.33\\
302.861593353245	0.33\\
302.861593353245	0.34\\
302.886045666052	0.34\\
302.886045666052	0.35\\
302.914056704836	0.35\\
302.914056704836	0.36\\
303.022350236134	0.36\\
303.022350236134	0.37\\
303.124703828432	0.37\\
303.124703828432	0.38\\
303.307685306915	0.38\\
303.307685306915	0.39\\
303.55356498009	0.39\\
303.55356498009	0.4\\
304.112371489593	0.4\\
304.112371489593	0.41\\
304.130539486197	0.41\\
304.130539486197	0.42\\
304.293261505343	0.42\\
304.293261505343	0.43\\
304.296171904544	0.43\\
304.296171904544	0.44\\
304.328442488989	0.44\\
304.328442488989	0.45\\
304.542880376911	0.45\\
304.542880376911	0.46\\
304.596684697582	0.46\\
304.596684697582	0.47\\
304.665299790599	0.47\\
304.665299790599	0.48\\
304.770400359408	0.48\\
304.770400359408	0.49\\
304.924566327551	0.49\\
304.924566327551	0.5\\
305.006553979887	0.5\\
305.006553979887	0.51\\
305.257278368573	0.51\\
305.257278368573	0.52\\
305.267105352128	0.52\\
305.267105352128	0.53\\
305.353064859643	0.53\\
305.353064859643	0.54\\
305.382414002099	0.54\\
305.382414002099	0.55\\
305.405730532005	0.55\\
305.405730532005	0.56\\
305.594596674614	0.56\\
305.594596674614	0.57\\
305.613420141213	0.57\\
305.613420141213	0.58\\
305.708469867278	0.58\\
305.708469867278	0.59\\
305.900635739701	0.59\\
305.900635739701	0.6\\
306.437534252083	0.6\\
306.437534252083	0.61\\
306.600568717548	0.61\\
306.600568717548	0.62\\
306.631164819339	0.62\\
306.631164819339	0.63\\
306.714699215195	0.63\\
306.714699215195	0.64\\
307.3162030131	0.64\\
307.3162030131	0.65\\
307.586808200513	0.65\\
307.586808200513	0.66\\
307.590644486675	0.66\\
307.590644486675	0.67\\
307.747827208645	0.67\\
307.747827208645	0.68\\
307.77402963041	0.68\\
307.77402963041	0.69\\
307.864720099515	0.69\\
307.864720099515	0.7\\
308.156196799683	0.7\\
308.156196799683	0.71\\
308.241719545741	0.71\\
308.241719545741	0.72\\
308.367654271822	0.72\\
308.367654271822	0.73\\
308.82614423562	0.73\\
308.82614423562	0.74\\
309.07222838171	0.74\\
309.07222838171	0.75\\
309.091519858238	0.75\\
309.091519858238	0.76\\
309.115343005971	0.76\\
309.115343005971	0.77\\
309.194788010214	0.77\\
309.194788010214	0.78\\
309.213176997623	0.78\\
309.213176997623	0.79\\
309.735815993813	0.79\\
309.735815993813	0.8\\
310.213289682217	0.8\\
310.213289682217	0.81\\
310.258485395437	0.81\\
310.258485395437	0.82\\
310.477186097341	0.82\\
310.477186097341	0.83\\
310.54366227933	0.83\\
310.54366227933	0.84\\
310.963395282413	0.84\\
310.963395282413	0.85\\
311.100900778469	0.85\\
311.100900778469	0.86\\
311.192767115415	0.86\\
311.192767115415	0.87\\
311.571706966405	0.87\\
311.571706966405	0.88\\
312.240514275039	0.88\\
312.240514275039	0.89\\
312.361319407942	0.89\\
312.361319407942	0.9\\
312.625808319554	0.9\\
312.625808319554	0.91\\
313.189158666023	0.91\\
313.189158666023	0.92\\
313.478539827352	0.92\\
313.478539827352	0.93\\
313.857216649462	0.93\\
313.857216649462	0.94\\
315.592050515107	0.94\\
315.592050515107	0.95\\
315.773888969087	0.95\\
315.773888969087	0.96\\
316.197644131985	0.96\\
316.197644131985	0.97\\
318.76457710822	0.97\\
318.76457710822	0.98\\
319.701731985428	0.98\\
319.701731985428	0.99\\
322.562895050769	0.99\\
322.562895050769	1\\
inf	1\\
};
\addlegendentry{ID $(\mathcal{L}\!=\!0.4$~y$)$}

\addplot [color=color2]
  table[row sep=crcr]{%
-inf	0\\
315.644630842145	0\\
315.644630842145	0.01\\
318.268744156928	0.01\\
318.268744156928	0.02\\
320.883363976604	0.02\\
320.883363976604	0.03\\
322.448367137951	0.03\\
322.448367137951	0.04\\
327.44460200309	0.04\\
327.44460200309	0.05\\
328.94753738536	0.05\\
328.94753738536	0.06\\
329.491002259874	0.06\\
329.491002259874	0.07\\
329.935969859963	0.07\\
329.935969859963	0.08\\
330.184484026235	0.08\\
330.184484026235	0.09\\
330.79321123251	0.09\\
330.79321123251	0.1\\
331.611047296001	0.1\\
331.611047296001	0.11\\
331.7100952485	0.11\\
331.7100952485	0.12\\
332.184927417679	0.12\\
332.184927417679	0.13\\
332.321176979732	0.13\\
332.321176979732	0.14\\
332.936736286841	0.14\\
332.936736286841	0.15\\
333.256871417038	0.15\\
333.256871417038	0.16\\
333.518270672172	0.16\\
333.518270672172	0.17\\
334.578373842346	0.17\\
334.578373842346	0.18\\
334.69907983099	0.18\\
334.69907983099	0.19\\
334.810025759401	0.19\\
334.810025759401	0.2\\
335.401964430641	0.2\\
335.401964430641	0.21\\
335.455177673362	0.21\\
335.455177673362	0.22\\
335.549468421437	0.22\\
335.549468421437	0.23\\
335.826421898798	0.23\\
335.826421898798	0.24\\
335.978762664118	0.24\\
335.978762664118	0.25\\
336.09551042192	0.25\\
336.09551042192	0.26\\
336.1648711135	0.26\\
336.1648711135	0.27\\
336.263214466928	0.27\\
336.263214466928	0.28\\
336.38269250249	0.28\\
336.38269250249	0.29\\
336.540675995534	0.29\\
336.540675995534	0.3\\
336.684962108447	0.3\\
336.684962108447	0.31\\
336.728684918371	0.31\\
336.728684918371	0.32\\
336.874898677799	0.32\\
336.874898677799	0.33\\
337.14597463091	0.33\\
337.14597463091	0.34\\
337.514149119779	0.34\\
337.514149119779	0.35\\
338.233431992734	0.35\\
338.233431992734	0.36\\
338.327138657922	0.36\\
338.327138657922	0.37\\
339.11672984266	0.37\\
339.11672984266	0.38\\
339.780471726849	0.38\\
339.780471726849	0.39\\
339.870978168848	0.39\\
339.870978168848	0.4\\
339.939216528542	0.4\\
339.939216528542	0.41\\
340.047761144357	0.41\\
340.047761144357	0.42\\
340.194130255689	0.42\\
340.194130255689	0.43\\
340.212428582226	0.43\\
340.212428582226	0.44\\
340.308204582965	0.44\\
340.308204582965	0.45\\
340.308926320687	0.45\\
340.308926320687	0.46\\
340.383684267051	0.46\\
340.383684267051	0.47\\
340.460751004449	0.47\\
340.460751004449	0.48\\
340.740503450965	0.48\\
340.740503450965	0.49\\
340.848225748614	0.49\\
340.848225748614	0.5\\
341.089086852437	0.5\\
341.089086852437	0.51\\
341.167906399327	0.51\\
341.167906399327	0.52\\
341.462006615132	0.52\\
341.462006615132	0.53\\
341.482042277306	0.53\\
341.482042277306	0.54\\
342.04868001149	0.54\\
342.04868001149	0.55\\
342.318875614262	0.55\\
342.318875614262	0.56\\
342.32794942737	0.56\\
342.32794942737	0.57\\
342.388617083287	0.57\\
342.388617083287	0.58\\
342.576936101325	0.58\\
342.576936101325	0.59\\
342.670294548751	0.59\\
342.670294548751	0.6\\
342.711983506514	0.6\\
342.711983506514	0.61\\
342.76245511559	0.61\\
342.76245511559	0.62\\
342.966638250311	0.62\\
342.966638250311	0.63\\
343.32511085553	0.63\\
343.32511085553	0.64\\
343.396171432266	0.64\\
343.396171432266	0.65\\
343.800349633804	0.65\\
343.800349633804	0.66\\
344.413903912204	0.66\\
344.413903912204	0.67\\
344.430566501045	0.67\\
344.430566501045	0.68\\
344.594530361353	0.68\\
344.594530361353	0.69\\
344.702675681592	0.69\\
344.702675681592	0.7\\
345.218046915329	0.7\\
345.218046915329	0.71\\
345.248535208585	0.71\\
345.248535208585	0.72\\
345.602746249944	0.72\\
345.602746249944	0.73\\
346.313556626082	0.73\\
346.313556626082	0.74\\
346.903111901405	0.74\\
346.903111901405	0.75\\
347.136784300131	0.75\\
347.136784300131	0.76\\
347.2826198259	0.76\\
347.2826198259	0.77\\
347.581148143948	0.77\\
347.581148143948	0.78\\
347.837216273057	0.78\\
347.837216273057	0.79\\
347.872933425523	0.79\\
347.872933425523	0.8\\
347.885631486668	0.8\\
347.885631486668	0.81\\
348.032773112993	0.81\\
348.032773112993	0.82\\
349.135766226547	0.82\\
349.135766226547	0.83\\
349.233359975701	0.83\\
349.233359975701	0.84\\
349.722034289003	0.84\\
349.722034289003	0.85\\
349.862278713418	0.85\\
349.862278713418	0.86\\
350.003912366519	0.86\\
350.003912366519	0.87\\
350.135308505746	0.87\\
350.135308505746	0.88\\
350.933545180324	0.88\\
350.933545180324	0.89\\
351.712798498295	0.89\\
351.712798498295	0.9\\
351.719470210155	0.9\\
351.719470210155	0.91\\
351.950414245692	0.91\\
351.950414245692	0.92\\
352.23933274074	0.92\\
352.23933274074	0.93\\
352.558793075821	0.93\\
352.558793075821	0.94\\
353.656255174169	0.94\\
353.656255174169	0.95\\
354.17796107697	0.95\\
354.17796107697	0.96\\
354.378474174657	0.96\\
354.378474174657	0.97\\
355.819597699955	0.97\\
355.819597699955	0.98\\
356.804717639536	0.98\\
356.804717639536	0.99\\
365.254605512353	0.99\\
365.254605512353	1\\
inf	1\\
};
\addlegendentry{$\theta\!=\!0.5\sigma (\mathcal{L}\!=\!0.43$~y$)$}

\addplot [color=color3]
  table[row sep=crcr]{%
-inf	0\\
342.513292591098	0\\
342.513292591098	0.01\\
344.08050217833	0.01\\
344.08050217833	0.02\\
351.317639872542	0.02\\
351.317639872542	0.03\\
351.563213947157	0.03\\
351.563213947157	0.04\\
352.460971989298	0.04\\
352.460971989298	0.05\\
354.931372882232	0.05\\
354.931372882232	0.06\\
355.015055786821	0.06\\
355.015055786821	0.07\\
355.70205673641	0.07\\
355.70205673641	0.08\\
356.153594066847	0.08\\
356.153594066847	0.09\\
356.189224043669	0.09\\
356.189224043669	0.1\\
357.365167239499	0.1\\
357.365167239499	0.11\\
358.059988231652	0.11\\
358.059988231652	0.12\\
358.461948901587	0.12\\
358.461948901587	0.13\\
358.487226919214	0.13\\
358.487226919214	0.14\\
358.692993596271	0.14\\
358.692993596271	0.15\\
358.737415793154	0.15\\
358.737415793154	0.16\\
359.61817494139	0.16\\
359.61817494139	0.17\\
359.661603614331	0.17\\
359.661603614331	0.18\\
359.806304167389	0.18\\
359.806304167389	0.19\\
359.906138239054	0.19\\
359.906138239054	0.2\\
360.051723156179	0.2\\
360.051723156179	0.21\\
360.178750457954	0.21\\
360.178750457954	0.22\\
360.968308969074	0.22\\
360.968308969074	0.23\\
361.278069144174	0.23\\
361.278069144174	0.24\\
361.285100741635	0.24\\
361.285100741635	0.25\\
361.407215054019	0.25\\
361.407215054019	0.26\\
361.981790345769	0.26\\
361.981790345769	0.27\\
362.403117889247	0.27\\
362.403117889247	0.28\\
362.975026002311	0.28\\
362.975026002311	0.29\\
363.037600164314	0.29\\
363.037600164314	0.3\\
363.263522191713	0.3\\
363.263522191713	0.31\\
363.73483683301	0.31\\
363.73483683301	0.32\\
363.940906746468	0.32\\
363.940906746468	0.33\\
363.952526396639	0.33\\
363.952526396639	0.34\\
364.036991732742	0.34\\
364.036991732742	0.35\\
364.11889944525	0.35\\
364.11889944525	0.36\\
364.51475112944	0.36\\
364.51475112944	0.37\\
364.756553948872	0.37\\
364.756553948872	0.38\\
364.836160472507	0.38\\
364.836160472507	0.39\\
365.425256353039	0.39\\
365.425256353039	0.4\\
366.016783552284	0.4\\
366.016783552284	0.41\\
366.298552955681	0.41\\
366.298552955681	0.42\\
366.299877300783	0.42\\
366.299877300783	0.43\\
366.818468029983	0.43\\
366.818468029983	0.44\\
367.049458076392	0.44\\
367.049458076392	0.45\\
367.254198186751	0.45\\
367.254198186751	0.46\\
367.408654126955	0.46\\
367.408654126955	0.47\\
367.44755987319	0.47\\
367.44755987319	0.48\\
367.512535122845	0.48\\
367.512535122845	0.49\\
368.233044420915	0.49\\
368.233044420915	0.5\\
368.661727528009	0.5\\
368.661727528009	0.51\\
369.076038329986	0.51\\
369.076038329986	0.52\\
369.154115505254	0.52\\
369.154115505254	0.53\\
369.15498771245	0.53\\
369.15498771245	0.54\\
369.17272557439	0.54\\
369.17272557439	0.55\\
369.528829058726	0.55\\
369.528829058726	0.56\\
369.54641904242	0.56\\
369.54641904242	0.57\\
369.611681734978	0.57\\
369.611681734978	0.58\\
369.77772725398	0.58\\
369.77772725398	0.59\\
369.921456596355	0.59\\
369.921456596355	0.6\\
370.386703857832	0.6\\
370.386703857832	0.61\\
370.641258512611	0.61\\
370.641258512611	0.62\\
370.883748360878	0.62\\
370.883748360878	0.63\\
370.955822127506	0.63\\
370.955822127506	0.64\\
371.118494253888	0.64\\
371.118494253888	0.65\\
371.293280133655	0.65\\
371.293280133655	0.66\\
371.443133741276	0.66\\
371.443133741276	0.67\\
371.556094882409	0.67\\
371.556094882409	0.68\\
372.180027472984	0.68\\
372.180027472984	0.69\\
372.420985556293	0.69\\
372.420985556293	0.7\\
372.690080876798	0.7\\
372.690080876798	0.71\\
373.239559984516	0.71\\
373.239559984516	0.72\\
374.810541343221	0.72\\
374.810541343221	0.73\\
375.715756528987	0.73\\
375.715756528987	0.74\\
376.030832962835	0.74\\
376.030832962835	0.75\\
376.045502576771	0.75\\
376.045502576771	0.76\\
376.267716966489	0.76\\
376.267716966489	0.77\\
376.288524096122	0.77\\
376.288524096122	0.78\\
377.113410575941	0.78\\
377.113410575941	0.79\\
377.615803039174	0.79\\
377.615803039174	0.8\\
377.984401285117	0.8\\
377.984401285117	0.81\\
378.226211717254	0.81\\
378.226211717254	0.82\\
378.25180872538	0.82\\
378.25180872538	0.83\\
378.516809436026	0.83\\
378.516809436026	0.84\\
378.846895764205	0.84\\
378.846895764205	0.85\\
378.909430516493	0.85\\
378.909430516493	0.86\\
378.994880283556	0.86\\
378.994880283556	0.87\\
379.089524276778	0.87\\
379.089524276778	0.88\\
379.323696826683	0.88\\
379.323696826683	0.89\\
379.357446479886	0.89\\
379.357446479886	0.9\\
379.55579230475	0.9\\
379.55579230475	0.91\\
380.253147136914	0.91\\
380.253147136914	0.92\\
380.823440366109	0.92\\
380.823440366109	0.93\\
382.170659124649	0.93\\
382.170659124649	0.94\\
382.596611562131	0.94\\
382.596611562131	0.95\\
384.142314933283	0.95\\
384.142314933283	0.96\\
385.717527105327	0.96\\
385.717527105327	0.97\\
386.871240873939	0.97\\
386.871240873939	0.98\\
390.481587544206	0.98\\
390.481587544206	0.99\\
391.649187545542	0.99\\
391.649187545542	1\\
inf	1\\
};
\addlegendentry{$\theta\!=\!\sigma (\mathcal{L}\!=\!0.47$~y$)$}

\addplot [color=color4]
  table[row sep=crcr]{%
-inf	0\\
368.032369628214	0\\
368.032369628214	0.01\\
373.340331789666	0.01\\
373.340331789666	0.02\\
373.66963228836	0.02\\
373.66963228836	0.03\\
373.692443626462	0.03\\
373.692443626462	0.04\\
373.886661259154	0.04\\
373.886661259154	0.05\\
374.256693051261	0.05\\
374.256693051261	0.06\\
375.737657081913	0.06\\
375.737657081913	0.07\\
378.29291326675	0.07\\
378.29291326675	0.08\\
379.687850443564	0.08\\
379.687850443564	0.09\\
380.029583263808	0.09\\
380.029583263808	0.1\\
381.215205073833	0.1\\
381.215205073833	0.11\\
381.750157660835	0.11\\
381.750157660835	0.12\\
382.028829726563	0.12\\
382.028829726563	0.13\\
382.157599458461	0.13\\
382.157599458461	0.14\\
383.052442357768	0.14\\
383.052442357768	0.15\\
383.519006325203	0.15\\
383.519006325203	0.16\\
383.533557365686	0.16\\
383.533557365686	0.17\\
384.019591110974	0.17\\
384.019591110974	0.18\\
384.16739910689	0.18\\
384.16739910689	0.19\\
384.555292123393	0.19\\
384.555292123393	0.2\\
385.212442784664	0.2\\
385.212442784664	0.21\\
385.342063652622	0.21\\
385.342063652622	0.22\\
386.048797826206	0.22\\
386.048797826206	0.23\\
386.095429388326	0.23\\
386.095429388326	0.24\\
386.641925866219	0.24\\
386.641925866219	0.25\\
386.803776875643	0.25\\
386.803776875643	0.26\\
386.928155399699	0.26\\
386.928155399699	0.27\\
387.247120610051	0.27\\
387.247120610051	0.28\\
387.273099681892	0.28\\
387.273099681892	0.29\\
387.861154766369	0.29\\
387.861154766369	0.3\\
388.102180774634	0.3\\
388.102180774634	0.31\\
388.145765688852	0.31\\
388.145765688852	0.32\\
388.194066926262	0.32\\
388.194066926262	0.33\\
388.515971532619	0.33\\
388.515971532619	0.34\\
388.723247639759	0.34\\
388.723247639759	0.35\\
388.756532085624	0.35\\
388.756532085624	0.36\\
388.812950980278	0.36\\
388.812950980278	0.37\\
388.909275912291	0.37\\
388.909275912291	0.38\\
389.122030153198	0.38\\
389.122030153198	0.39\\
389.146475547893	0.39\\
389.146475547893	0.4\\
389.288362437789	0.4\\
389.288362437789	0.41\\
389.470093874737	0.41\\
389.470093874737	0.42\\
389.567709588825	0.42\\
389.567709588825	0.43\\
389.624063298864	0.43\\
389.624063298864	0.44\\
389.753584477194	0.44\\
389.753584477194	0.45\\
390.415923697636	0.45\\
390.415923697636	0.46\\
390.859998281422	0.46\\
390.859998281422	0.47\\
391.238843767652	0.47\\
391.238843767652	0.48\\
391.316946115988	0.48\\
391.316946115988	0.49\\
391.739339046312	0.49\\
391.739339046312	0.5\\
392.00774002921	0.5\\
392.00774002921	0.51\\
392.098971070109	0.51\\
392.098971070109	0.52\\
392.380829389972	0.52\\
392.380829389972	0.53\\
392.715517724105	0.53\\
392.715517724105	0.54\\
393.049398428894	0.54\\
393.049398428894	0.55\\
393.526556656909	0.55\\
393.526556656909	0.56\\
393.909679462378	0.56\\
393.909679462378	0.57\\
394.58070227295	0.57\\
394.58070227295	0.58\\
395.29151053623	0.58\\
395.29151053623	0.59\\
395.942448178463	0.59\\
395.942448178463	0.6\\
396.052888366994	0.6\\
396.052888366994	0.61\\
396.604407699928	0.61\\
396.604407699928	0.62\\
396.835026434815	0.62\\
396.835026434815	0.63\\
397.200764064673	0.63\\
397.200764064673	0.64\\
397.422678024128	0.64\\
397.422678024128	0.65\\
398.102604312367	0.65\\
398.102604312367	0.66\\
399.175087527923	0.66\\
399.175087527923	0.67\\
399.370737210234	0.67\\
399.370737210234	0.68\\
399.549628439875	0.68\\
399.549628439875	0.69\\
399.759187146087	0.69\\
399.759187146087	0.7\\
400.298531468743	0.7\\
400.298531468743	0.71\\
400.416452086747	0.71\\
400.416452086747	0.72\\
400.42032758927	0.72\\
400.42032758927	0.73\\
400.76662762208	0.73\\
400.76662762208	0.74\\
402.010148914351	0.74\\
402.010148914351	0.75\\
402.172771326925	0.75\\
402.172771326925	0.76\\
402.254483639892	0.76\\
402.254483639892	0.77\\
402.260857924326	0.77\\
402.260857924326	0.78\\
402.560680516699	0.78\\
402.560680516699	0.79\\
402.664397889565	0.79\\
402.664397889565	0.8\\
402.740059345105	0.8\\
402.740059345105	0.81\\
404.870423108617	0.81\\
404.870423108617	0.82\\
405.026962712419	0.82\\
405.026962712419	0.83\\
405.048887792467	0.83\\
405.048887792467	0.84\\
405.173658944358	0.84\\
405.173658944358	0.85\\
405.445296951057	0.85\\
405.445296951057	0.86\\
405.48605670323	0.86\\
405.48605670323	0.87\\
406.924190122937	0.87\\
406.924190122937	0.88\\
407.028335448158	0.88\\
407.028335448158	0.89\\
407.491308647812	0.89\\
407.491308647812	0.9\\
408.232875613349	0.9\\
408.232875613349	0.91\\
408.888271964392	0.91\\
408.888271964392	0.92\\
409.256903169525	0.92\\
409.256903169525	0.93\\
410.154452858449	0.93\\
410.154452858449	0.94\\
410.427767495334	0.94\\
410.427767495334	0.95\\
411.552845532084	0.95\\
411.552845532084	0.96\\
413.605247035875	0.96\\
413.605247035875	0.97\\
415.251742206548	0.97\\
415.251742206548	0.98\\
418.915259476009	0.98\\
418.915259476009	0.99\\
419.349156750995	0.99\\
419.349156750995	1\\
inf	1\\
};
\addlegendentry{$\theta\!=\!1.5\sigma (\mathcal{L}\!=\!0.52$~y$)$}

\addplot [color=color5]
  table[row sep=crcr]{%
-inf	0\\
395.714541219134	0\\
395.714541219134	0.01\\
398.212210840538	0.01\\
398.212210840538	0.02\\
398.906174759937	0.02\\
398.906174759937	0.03\\
400.382841474552	0.03\\
400.382841474552	0.04\\
407.304138401349	0.04\\
407.304138401349	0.05\\
407.524124250139	0.05\\
407.524124250139	0.06\\
408.084303767274	0.06\\
408.084303767274	0.07\\
408.594539147411	0.07\\
408.594539147411	0.08\\
409.16052445754	0.08\\
409.16052445754	0.09\\
409.453604565156	0.09\\
409.453604565156	0.1\\
410.785668438337	0.1\\
410.785668438337	0.11\\
412.732792395732	0.11\\
412.732792395732	0.12\\
413.418029049969	0.12\\
413.418029049969	0.13\\
414.315319235661	0.13\\
414.315319235661	0.14\\
414.491929053971	0.14\\
414.491929053971	0.15\\
415.492729547226	0.15\\
415.492729547226	0.16\\
416.154796763927	0.16\\
416.154796763927	0.17\\
416.164718227305	0.17\\
416.164718227305	0.18\\
416.203576437271	0.18\\
416.203576437271	0.19\\
416.565750481551	0.19\\
416.565750481551	0.2\\
416.613497192613	0.2\\
416.613497192613	0.21\\
417.166822575892	0.21\\
417.166822575892	0.22\\
417.223879115456	0.22\\
417.223879115456	0.23\\
417.675640898014	0.23\\
417.675640898014	0.24\\
417.939522468616	0.24\\
417.939522468616	0.25\\
418.067539293836	0.25\\
418.067539293836	0.26\\
418.221508970447	0.26\\
418.221508970447	0.27\\
418.258865303947	0.27\\
418.258865303947	0.28\\
418.501814883901	0.28\\
418.501814883901	0.29\\
419.091016376332	0.29\\
419.091016376332	0.3\\
419.141517180261	0.3\\
419.141517180261	0.31\\
419.193566528181	0.31\\
419.193566528181	0.32\\
419.27775904891	0.32\\
419.27775904891	0.33\\
419.507086743571	0.33\\
419.507086743571	0.34\\
419.55952027731	0.34\\
419.55952027731	0.35\\
420.079314795075	0.35\\
420.079314795075	0.36\\
420.411155408232	0.36\\
420.411155408232	0.37\\
420.545381583279	0.37\\
420.545381583279	0.38\\
420.621212558605	0.38\\
420.621212558605	0.39\\
420.823208320244	0.39\\
420.823208320244	0.4\\
420.858636421354	0.4\\
420.858636421354	0.41\\
420.967924063916	0.41\\
420.967924063916	0.42\\
421.074341336389	0.42\\
421.074341336389	0.43\\
421.10212504562	0.43\\
421.10212504562	0.44\\
421.254467779335	0.44\\
421.254467779335	0.45\\
422.389936405584	0.45\\
422.389936405584	0.46\\
422.648380894863	0.46\\
422.648380894863	0.47\\
423.055816352602	0.47\\
423.055816352602	0.48\\
424.257701384775	0.48\\
424.257701384775	0.49\\
424.445619608581	0.49\\
424.445619608581	0.5\\
424.503629927136	0.5\\
424.503629927136	0.51\\
424.632059644628	0.51\\
424.632059644628	0.52\\
425.396335272015	0.52\\
425.396335272015	0.53\\
425.57172604252	0.53\\
425.57172604252	0.54\\
426.158156771852	0.54\\
426.158156771852	0.55\\
426.542270339071	0.55\\
426.542270339071	0.56\\
426.594259911725	0.56\\
426.594259911725	0.57\\
426.726274379144	0.57\\
426.726274379144	0.58\\
427.366207718676	0.58\\
427.366207718676	0.59\\
427.483442509119	0.59\\
427.483442509119	0.6\\
427.970169060522	0.6\\
427.970169060522	0.61\\
428.128759295118	0.61\\
428.128759295118	0.62\\
428.585820910622	0.62\\
428.585820910622	0.63\\
428.861463262262	0.63\\
428.861463262262	0.64\\
429.248337572751	0.64\\
429.248337572751	0.65\\
429.585526205909	0.65\\
429.585526205909	0.66\\
429.893790230649	0.66\\
429.893790230649	0.67\\
430.024413738385	0.67\\
430.024413738385	0.68\\
430.1510879202	0.68\\
430.1510879202	0.69\\
430.318448822304	0.69\\
430.318448822304	0.7\\
430.555053639032	0.7\\
430.555053639032	0.71\\
430.562058124567	0.71\\
430.562058124567	0.72\\
431.26461194258	0.72\\
431.26461194258	0.73\\
431.311195557855	0.73\\
431.311195557855	0.74\\
431.350146930635	0.74\\
431.350146930635	0.75\\
431.58840305343	0.75\\
431.58840305343	0.76\\
432.234129120196	0.76\\
432.234129120196	0.77\\
432.237547214752	0.77\\
432.237547214752	0.78\\
432.263449524151	0.78\\
432.263449524151	0.79\\
432.983104371074	0.79\\
432.983104371074	0.8\\
433.346508952554	0.8\\
433.346508952554	0.81\\
433.825088313252	0.81\\
433.825088313252	0.82\\
433.903246484733	0.82\\
433.903246484733	0.83\\
434.668481351113	0.83\\
434.668481351113	0.84\\
435.876723833918	0.84\\
435.876723833918	0.85\\
436.026497584441	0.85\\
436.026497584441	0.86\\
436.346578408516	0.86\\
436.346578408516	0.87\\
436.374696252859	0.87\\
436.374696252859	0.88\\
436.376579637159	0.88\\
436.376579637159	0.89\\
436.73836816631	0.89\\
436.73836816631	0.9\\
437.129131434639	0.9\\
437.129131434639	0.91\\
437.546274533456	0.91\\
437.546274533456	0.92\\
437.924055802514	0.92\\
437.924055802514	0.93\\
440.957438705588	0.93\\
440.957438705588	0.94\\
442.762341782477	0.94\\
442.762341782477	0.95\\
443.286893694457	0.95\\
443.286893694457	0.96\\
443.362303207686	0.96\\
443.362303207686	0.97\\
444.619635374684	0.97\\
444.619635374684	0.98\\
446.526292224734	0.98\\
446.526292224734	0.99\\
447.14474845043	0.99\\
447.14474845043	1\\
inf	1\\
};
\addlegendentry{$\theta\!=\!2\sigma (\mathcal{L}\!=\!0.57$~y$)$}

\addplot [color=color6]
  table[row sep=crcr]{%
-inf	0\\
424.919961193197	0\\
424.919961193197	0.01\\
436.217122268115	0.01\\
436.217122268115	0.02\\
436.818377552821	0.02\\
436.818377552821	0.03\\
438.624941250819	0.03\\
438.624941250819	0.04\\
438.925012903528	0.04\\
438.925012903528	0.05\\
440.765635921846	0.05\\
440.765635921846	0.06\\
441.21716546292	0.06\\
441.21716546292	0.07\\
441.638668472384	0.07\\
441.638668472384	0.08\\
442.331659464594	0.08\\
442.331659464594	0.09\\
443.893189304928	0.09\\
443.893189304928	0.1\\
444.77161143097	0.1\\
444.77161143097	0.11\\
445.193933489202	0.11\\
445.193933489202	0.12\\
445.501124374891	0.12\\
445.501124374891	0.13\\
446.321474241704	0.13\\
446.321474241704	0.14\\
447.387406841101	0.14\\
447.387406841101	0.15\\
447.740118497289	0.15\\
447.740118497289	0.16\\
448.074617929303	0.16\\
448.074617929303	0.17\\
448.476825507815	0.17\\
448.476825507815	0.18\\
448.661321224427	0.18\\
448.661321224427	0.19\\
449.09961656332	0.19\\
449.09961656332	0.2\\
449.289445161446	0.2\\
449.289445161446	0.21\\
449.567920361706	0.21\\
449.567920361706	0.22\\
450.825497051784	0.22\\
450.825497051784	0.23\\
451.677014311442	0.23\\
451.677014311442	0.24\\
452.111355926548	0.24\\
452.111355926548	0.25\\
453.039156126795	0.25\\
453.039156126795	0.26\\
453.283460082182	0.26\\
453.283460082182	0.27\\
453.508081781097	0.27\\
453.508081781097	0.28\\
453.558480682932	0.28\\
453.558480682932	0.29\\
453.631289626033	0.29\\
453.631289626033	0.3\\
454.688776836379	0.3\\
454.688776836379	0.31\\
454.755009571562	0.31\\
454.755009571562	0.32\\
455.032532840066	0.32\\
455.032532840066	0.33\\
455.530353554188	0.33\\
455.530353554188	0.34\\
455.993062726956	0.34\\
455.993062726956	0.35\\
456.401395413519	0.35\\
456.401395413519	0.36\\
456.551166798679	0.36\\
456.551166798679	0.37\\
456.590179706451	0.37\\
456.590179706451	0.38\\
456.991243302049	0.38\\
456.991243302049	0.39\\
457.393271869348	0.39\\
457.393271869348	0.4\\
457.595239256631	0.4\\
457.595239256631	0.41\\
457.691763650108	0.41\\
457.691763650108	0.42\\
459.028801559856	0.42\\
459.028801559856	0.43\\
459.177545537112	0.43\\
459.177545537112	0.44\\
459.281442403269	0.44\\
459.281442403269	0.45\\
459.426298711297	0.45\\
459.426298711297	0.46\\
459.431504067486	0.46\\
459.431504067486	0.47\\
459.473925402024	0.47\\
459.473925402024	0.48\\
460.522267692761	0.48\\
460.522267692761	0.49\\
460.812233568447	0.49\\
460.812233568447	0.5\\
460.914509467002	0.5\\
460.914509467002	0.51\\
461.131974825485	0.51\\
461.131974825485	0.52\\
461.231763958537	0.52\\
461.231763958537	0.53\\
461.48520169854	0.53\\
461.48520169854	0.54\\
461.633770742299	0.54\\
461.633770742299	0.55\\
462.101414214264	0.55\\
462.101414214264	0.56\\
462.695295601665	0.56\\
462.695295601665	0.57\\
463.279284913471	0.57\\
463.279284913471	0.58\\
463.307582303786	0.58\\
463.307582303786	0.59\\
463.35662107069	0.59\\
463.35662107069	0.6\\
463.779531006011	0.6\\
463.779531006011	0.61\\
463.867122670963	0.61\\
463.867122670963	0.62\\
464.55702839944	0.62\\
464.55702839944	0.63\\
464.9005870349	0.63\\
464.9005870349	0.64\\
464.918198564397	0.64\\
464.918198564397	0.65\\
465.139181007276	0.65\\
465.139181007276	0.66\\
465.459332677642	0.66\\
465.459332677642	0.67\\
465.474375205767	0.67\\
465.474375205767	0.68\\
465.627343636502	0.68\\
465.627343636502	0.69\\
465.74453295133	0.69\\
465.74453295133	0.7\\
466.553573896806	0.7\\
466.553573896806	0.71\\
466.581359590326	0.71\\
466.581359590326	0.72\\
466.822764684793	0.72\\
466.822764684793	0.73\\
467.21992216351	0.73\\
467.21992216351	0.74\\
467.622843616538	0.74\\
467.622843616538	0.75\\
468.865974477293	0.75\\
468.865974477293	0.76\\
469.384485773471	0.76\\
469.384485773471	0.77\\
470.303078992866	0.77\\
470.303078992866	0.78\\
470.324090638426	0.78\\
470.324090638426	0.79\\
470.99081937426	0.79\\
470.99081937426	0.8\\
472.139361858547	0.8\\
472.139361858547	0.81\\
472.147325736759	0.81\\
472.147325736759	0.82\\
472.776536668267	0.82\\
472.776536668267	0.83\\
473.588118496503	0.83\\
473.588118496503	0.84\\
473.879074277374	0.84\\
473.879074277374	0.85\\
473.980741560623	0.85\\
473.980741560623	0.86\\
474.243649455373	0.86\\
474.243649455373	0.87\\
474.261533757862	0.87\\
474.261533757862	0.88\\
474.652309523827	0.88\\
474.652309523827	0.89\\
475.116993598658	0.89\\
475.116993598658	0.9\\
475.488399309087	0.9\\
475.488399309087	0.91\\
476.001733137463	0.91\\
476.001733137463	0.92\\
477.401229636086	0.92\\
477.401229636086	0.93\\
477.704766949702	0.93\\
477.704766949702	0.94\\
478.165697936133	0.94\\
478.165697936133	0.95\\
479.085967219685	0.95\\
479.085967219685	0.96\\
482.372820909724	0.96\\
482.372820909724	0.97\\
484.063215959916	0.97\\
484.063215959916	0.98\\
486.302147070454	0.98\\
486.302147070454	0.99\\
487.229409473151	0.99\\
487.229409473151	1\\
inf	1\\
};
\addlegendentry{$\theta\!=\!2.5\sigma (\mathcal{L}\!=\!0.62$~y$)$}

\addplot [color=color7]
  table[row sep=crcr]{%
-inf	0\\
459.265355784083	0\\
459.265355784083	0.01\\
463.292328511442	0.01\\
463.292328511442	0.02\\
465.718420940326	0.02\\
465.718420940326	0.03\\
471.516702756495	0.03\\
471.516702756495	0.04\\
472.336427305791	0.04\\
472.336427305791	0.05\\
473.988128989524	0.05\\
473.988128989524	0.06\\
476.795936856527	0.06\\
476.795936856527	0.07\\
478.046262102475	0.07\\
478.046262102475	0.08\\
478.277678658172	0.08\\
478.277678658172	0.09\\
479.092801565343	0.09\\
479.092801565343	0.1\\
480.222858498962	0.1\\
480.222858498962	0.11\\
480.515745300332	0.11\\
480.515745300332	0.12\\
481.032448636621	0.12\\
481.032448636621	0.13\\
481.290681864095	0.13\\
481.290681864095	0.14\\
481.372774802044	0.14\\
481.372774802044	0.15\\
481.392169460355	0.15\\
481.392169460355	0.16\\
481.772205272091	0.16\\
481.772205272091	0.17\\
482.030469688665	0.17\\
482.030469688665	0.18\\
483.55516151096	0.18\\
483.55516151096	0.19\\
484.39431243027	0.19\\
484.39431243027	0.2\\
485.510600229209	0.2\\
485.510600229209	0.21\\
485.877028772779	0.21\\
485.877028772779	0.22\\
486.678645937059	0.22\\
486.678645937059	0.23\\
486.89186231835	0.23\\
486.89186231835	0.24\\
486.957413638814	0.24\\
486.957413638814	0.25\\
487.018183740404	0.25\\
487.018183740404	0.26\\
487.837710614956	0.26\\
487.837710614956	0.27\\
488.095245140328	0.27\\
488.095245140328	0.28\\
488.27852928677	0.28\\
488.27852928677	0.29\\
489.436855708245	0.29\\
489.436855708245	0.3\\
490.025304844241	0.3\\
490.025304844241	0.31\\
490.25822349389	0.31\\
490.25822349389	0.32\\
490.414119196499	0.32\\
490.414119196499	0.33\\
490.830540449144	0.33\\
490.830540449144	0.34\\
491.155515186257	0.34\\
491.155515186257	0.35\\
491.308896935635	0.35\\
491.308896935635	0.36\\
491.447567194874	0.36\\
491.447567194874	0.37\\
491.961304187668	0.37\\
491.961304187668	0.38\\
492.452904412498	0.38\\
492.452904412498	0.39\\
492.951863423142	0.39\\
492.951863423142	0.4\\
493.502761585454	0.4\\
493.502761585454	0.41\\
493.51332509582	0.41\\
493.51332509582	0.42\\
493.880248849145	0.42\\
493.880248849145	0.43\\
494.093898248908	0.43\\
494.093898248908	0.44\\
494.211843560319	0.44\\
494.211843560319	0.45\\
494.281385750076	0.45\\
494.281385750076	0.46\\
494.29329384485	0.46\\
494.29329384485	0.47\\
496.196873272802	0.47\\
496.196873272802	0.48\\
496.40007159688	0.48\\
496.40007159688	0.49\\
496.976762065006	0.49\\
496.976762065006	0.5\\
498.19866061018	0.5\\
498.19866061018	0.51\\
498.595434309924	0.51\\
498.595434309924	0.52\\
498.643480724399	0.52\\
498.643480724399	0.53\\
498.714367597545	0.53\\
498.714367597545	0.54\\
499.193866220207	0.54\\
499.193866220207	0.55\\
499.273792485837	0.55\\
499.273792485837	0.56\\
499.640466734376	0.56\\
499.640466734376	0.57\\
499.910275878018	0.57\\
499.910275878018	0.58\\
499.998002710555	0.58\\
499.998002710555	0.59\\
500.54351467638	0.59\\
500.54351467638	0.6\\
501.003883843795	0.6\\
501.003883843795	0.61\\
501.050567700107	0.61\\
501.050567700107	0.62\\
501.36891014475	0.62\\
501.36891014475	0.63\\
501.405772835894	0.63\\
501.405772835894	0.64\\
501.41480924205	0.64\\
501.41480924205	0.65\\
501.793932860997	0.65\\
501.793932860997	0.66\\
501.849925204637	0.66\\
501.849925204637	0.67\\
501.89741381588	0.67\\
501.89741381588	0.68\\
502.065613538453	0.68\\
502.065613538453	0.69\\
502.845838684627	0.69\\
502.845838684627	0.7\\
502.852983207435	0.7\\
502.852983207435	0.71\\
502.962228013927	0.71\\
502.962228013927	0.72\\
503.525535741292	0.72\\
503.525535741292	0.73\\
503.605055171036	0.73\\
503.605055171036	0.74\\
503.792126370988	0.74\\
503.792126370988	0.75\\
504.108689467234	0.75\\
504.108689467234	0.76\\
505.040869500563	0.76\\
505.040869500563	0.77\\
505.917607953345	0.77\\
505.917607953345	0.78\\
505.957540502359	0.78\\
505.957540502359	0.79\\
506.709882527715	0.79\\
506.709882527715	0.8\\
507.726206093305	0.8\\
507.726206093305	0.81\\
507.923751134509	0.81\\
507.923751134509	0.82\\
508.502350235789	0.82\\
508.502350235789	0.83\\
508.768816234167	0.83\\
508.768816234167	0.84\\
509.96015536099	0.84\\
509.96015536099	0.85\\
510.521982608469	0.85\\
510.521982608469	0.86\\
510.791826520745	0.86\\
510.791826520745	0.87\\
511.095834695335	0.87\\
511.095834695335	0.88\\
512.291451123764	0.88\\
512.291451123764	0.89\\
512.883371753289	0.89\\
512.883371753289	0.9\\
513.156495873136	0.9\\
513.156495873136	0.91\\
513.281204238576	0.91\\
513.281204238576	0.92\\
513.846185711644	0.92\\
513.846185711644	0.93\\
514.564153654915	0.93\\
514.564153654915	0.94\\
516.134451960647	0.94\\
516.134451960647	0.95\\
516.664614825253	0.95\\
516.664614825253	0.96\\
522.220936108575	0.96\\
522.220936108575	0.97\\
524.157783537164	0.97\\
524.157783537164	0.98\\
528.008677287949	0.98\\
528.008677287949	0.99\\
534.386951198124	0.99\\
534.386951198124	1\\
inf	1\\
};
\addlegendentry{$\theta\!=\!3\sigma (\mathcal{L}\!=\!0.67$~y$)$}

\end{axis}

\end{tikzpicture}%

%% file: Images/Scenario_2_cdf.tex
%
%
\begin{tikzpicture}
\begin{axis}[%
width=\cwidth,
height=\cheight,
at={(0.758in,0.481in)},
scale only axis,
unbounded coords=jump,
xmin=0,
xmax=600,
xlabel style={font=\color{white!15!black}},
xlabel={MSE},
ymin=0,
ymax=1,
ylabel style={font=\color{white!15!black}},
ylabel={Empirical CDF},
axis background/.style={fill=white},
xmajorgrids,
ymajorgrids,
legend style={at={(0.005,0.99)}, anchor=north west, legend cell align=left, align=left, draw=white!15!black,font={\footnotesize}}
]
\addplot [color=color1]
  table[row sep=crcr]{%
-inf	0\\
306.20236618389	0\\
306.20236618389	0.01\\
312.358513773967	0.01\\
312.358513773967	0.02\\
313.059783849609	0.02\\
313.059783849609	0.03\\
314.528161078401	0.03\\
314.528161078401	0.04\\
317.211369115151	0.04\\
317.211369115151	0.05\\
317.53849186091	0.05\\
317.53849186091	0.06\\
317.612329108538	0.06\\
317.612329108538	0.07\\
317.688571590471	0.07\\
317.688571590471	0.08\\
318.015849177707	0.08\\
318.015849177707	0.09\\
318.426055336551	0.09\\
318.426055336551	0.1\\
318.788526970668	0.1\\
318.788526970668	0.11\\
319.135840087491	0.11\\
319.135840087491	0.12\\
319.343588193551	0.12\\
319.343588193551	0.13\\
319.715100093893	0.13\\
319.715100093893	0.14\\
319.934795121675	0.14\\
319.934795121675	0.15\\
320.363492873701	0.15\\
320.363492873701	0.16\\
321.212709909882	0.16\\
321.212709909882	0.17\\
321.231076638496	0.17\\
321.231076638496	0.18\\
321.376670246037	0.18\\
321.376670246037	0.19\\
321.814194683358	0.19\\
321.814194683358	0.2\\
321.816953275959	0.2\\
321.816953275959	0.21\\
321.94403248753	0.21\\
321.94403248753	0.22\\
322.024542092712	0.22\\
322.024542092712	0.23\\
322.264742253632	0.23\\
322.264742253632	0.24\\
322.456385712201	0.24\\
322.456385712201	0.25\\
322.817156308235	0.25\\
322.817156308235	0.26\\
322.978891110685	0.26\\
322.978891110685	0.27\\
323.054311752419	0.27\\
323.054311752419	0.28\\
323.105653870789	0.28\\
323.105653870789	0.29\\
323.149247949733	0.29\\
323.149247949733	0.3\\
323.179541582229	0.3\\
323.179541582229	0.31\\
323.202762174878	0.31\\
323.202762174878	0.32\\
323.203143530733	0.32\\
323.203143530733	0.33\\
323.575470444276	0.33\\
323.575470444276	0.34\\
323.869266715178	0.34\\
323.869266715178	0.35\\
324.938069464802	0.35\\
324.938069464802	0.36\\
325.029576562438	0.36\\
325.029576562438	0.37\\
325.063701888583	0.37\\
325.063701888583	0.38\\
325.254662575543	0.38\\
325.254662575543	0.39\\
325.316597927063	0.39\\
325.316597927063	0.4\\
325.521023565645	0.4\\
325.521023565645	0.41\\
325.536059638864	0.41\\
325.536059638864	0.42\\
325.567178634919	0.42\\
325.567178634919	0.43\\
325.66361657168	0.43\\
325.66361657168	0.44\\
325.793379610761	0.44\\
325.793379610761	0.45\\
326.287182579798	0.45\\
326.287182579798	0.46\\
326.304274674649	0.46\\
326.304274674649	0.47\\
326.69897936095	0.47\\
326.69897936095	0.48\\
327.06079426552	0.48\\
327.06079426552	0.49\\
327.1186897653	0.49\\
327.1186897653	0.5\\
327.18617349246	0.5\\
327.18617349246	0.51\\
327.317215570806	0.51\\
327.317215570806	0.52\\
327.366391442309	0.52\\
327.366391442309	0.53\\
327.409406550267	0.53\\
327.409406550267	0.54\\
327.428817652346	0.54\\
327.428817652346	0.55\\
327.584438124592	0.55\\
327.584438124592	0.56\\
327.720809173434	0.56\\
327.720809173434	0.57\\
327.7842784197	0.57\\
327.7842784197	0.58\\
327.86282250919	0.58\\
327.86282250919	0.59\\
328.071616962383	0.59\\
328.071616962383	0.6\\
328.257006936949	0.6\\
328.257006936949	0.61\\
328.342510518986	0.61\\
328.342510518986	0.62\\
328.361258455513	0.62\\
328.361258455513	0.63\\
328.481112906426	0.63\\
328.481112906426	0.64\\
328.631881579638	0.64\\
328.631881579638	0.65\\
329.055508768489	0.65\\
329.055508768489	0.66\\
329.057173150868	0.66\\
329.057173150868	0.67\\
329.200652652183	0.67\\
329.200652652183	0.68\\
329.264352291194	0.68\\
329.264352291194	0.69\\
329.329014246193	0.69\\
329.329014246193	0.7\\
329.362197536661	0.7\\
329.362197536661	0.71\\
329.562853596516	0.71\\
329.562853596516	0.72\\
329.779471362315	0.72\\
329.779471362315	0.73\\
329.83043935634	0.73\\
329.83043935634	0.74\\
330.298478876648	0.74\\
330.298478876648	0.75\\
330.399706441826	0.75\\
330.399706441826	0.76\\
330.417593830191	0.76\\
330.417593830191	0.77\\
331.405791543525	0.77\\
331.405791543525	0.78\\
331.666370765195	0.78\\
331.666370765195	0.79\\
331.684183954614	0.79\\
331.684183954614	0.8\\
332.047594097709	0.8\\
332.047594097709	0.81\\
332.313280947035	0.81\\
332.313280947035	0.82\\
332.87197394301	0.82\\
332.87197394301	0.83\\
332.914009191508	0.83\\
332.914009191508	0.84\\
333.152552302851	0.84\\
333.152552302851	0.85\\
333.207095842009	0.85\\
333.207095842009	0.86\\
333.366266424923	0.86\\
333.366266424923	0.87\\
333.94665476322	0.87\\
333.94665476322	0.88\\
334.559250011967	0.88\\
334.559250011967	0.89\\
335.44494705272	0.89\\
335.44494705272	0.9\\
336.544157256333	0.9\\
336.544157256333	0.91\\
336.837403508233	0.91\\
336.837403508233	0.92\\
337.773242481808	0.92\\
337.773242481808	0.93\\
338.435354951783	0.93\\
338.435354951783	0.94\\
338.469962811616	0.94\\
338.469962811616	0.95\\
338.85204904316	0.95\\
338.85204904316	0.96\\
339.04779776523	0.96\\
339.04779776523	0.97\\
341.429219630815	0.97\\
341.429219630815	0.98\\
342.941227303339	0.98\\
342.941227303339	0.99\\
343.119991791328	0.99\\
343.119991791328	1\\
inf	1\\
};
\addlegendentry{ID $(\mathcal{L}\!=\!0.4$~y$)$}
\addplot [color=color2]
  table[row sep=crcr]{%
-inf	0\\
322.77296830224	0\\
322.77296830224	0.01\\
325.156197025795	0.01\\
325.156197025795	0.02\\
325.18889308989	0.02\\
325.18889308989	0.03\\
327.086488186604	0.03\\
327.086488186604	0.04\\
327.437857839492	0.04\\
327.437857839492	0.05\\
328.136009035653	0.05\\
328.136009035653	0.06\\
328.295935178642	0.06\\
328.295935178642	0.07\\
328.822075257394	0.07\\
328.822075257394	0.08\\
328.859235976396	0.08\\
328.859235976396	0.09\\
329.523549701599	0.09\\
329.523549701599	0.1\\
329.620202410233	0.1\\
329.620202410233	0.11\\
329.622652845721	0.11\\
329.622652845721	0.12\\
329.664800265909	0.12\\
329.664800265909	0.13\\
329.70953513086	0.13\\
329.70953513086	0.14\\
329.722724967644	0.14\\
329.722724967644	0.15\\
330.389833706517	0.15\\
330.389833706517	0.16\\
330.688278744906	0.16\\
330.688278744906	0.17\\
330.916857222557	0.17\\
330.916857222557	0.18\\
330.974502254552	0.18\\
330.974502254552	0.19\\
331.950584545717	0.19\\
331.950584545717	0.2\\
331.996512811846	0.2\\
331.996512811846	0.21\\
332.551818247165	0.21\\
332.551818247165	0.22\\
332.599567553893	0.22\\
332.599567553893	0.23\\
332.982004337206	0.23\\
332.982004337206	0.24\\
333.072575805221	0.24\\
333.072575805221	0.25\\
333.942279613518	0.25\\
333.942279613518	0.26\\
333.984331766252	0.26\\
333.984331766252	0.27\\
334.075644900579	0.27\\
334.075644900579	0.28\\
334.371150391464	0.28\\
334.371150391464	0.29\\
334.435221856777	0.29\\
334.435221856777	0.3\\
334.493396021052	0.3\\
334.493396021052	0.31\\
334.996796875743	0.31\\
334.996796875743	0.32\\
335.120877945029	0.32\\
335.120877945029	0.33\\
335.490873157612	0.33\\
335.490873157612	0.34\\
335.616239428934	0.34\\
335.616239428934	0.35\\
335.627391883249	0.35\\
335.627391883249	0.36\\
335.755547817783	0.36\\
335.755547817783	0.37\\
335.845521681431	0.37\\
335.845521681431	0.38\\
335.950552229907	0.38\\
335.950552229907	0.39\\
335.978750172249	0.39\\
335.978750172249	0.4\\
336.347415529577	0.4\\
336.347415529577	0.41\\
337.296991643593	0.41\\
337.296991643593	0.42\\
337.577662613042	0.42\\
337.577662613042	0.43\\
337.602774100933	0.43\\
337.602774100933	0.44\\
337.754260141841	0.44\\
337.754260141841	0.45\\
337.917673118714	0.45\\
337.917673118714	0.46\\
338.205534612477	0.46\\
338.205534612477	0.47\\
338.541970255937	0.47\\
338.541970255937	0.48\\
338.566667824655	0.48\\
338.566667824655	0.49\\
338.681630311715	0.49\\
338.681630311715	0.5\\
338.907588946983	0.5\\
338.907588946983	0.51\\
339.102426923472	0.51\\
339.102426923472	0.52\\
339.132926528143	0.52\\
339.132926528143	0.53\\
339.254068574773	0.53\\
339.254068574773	0.54\\
339.28868567489	0.54\\
339.28868567489	0.55\\
339.35765667816	0.55\\
339.35765667816	0.56\\
339.400017681516	0.56\\
339.400017681516	0.57\\
339.564410386319	0.57\\
339.564410386319	0.58\\
339.790052668637	0.58\\
339.790052668637	0.59\\
340.397849539845	0.59\\
340.397849539845	0.6\\
340.454889679299	0.6\\
340.454889679299	0.61\\
340.606909062882	0.61\\
340.606909062882	0.62\\
340.885731691332	0.62\\
340.885731691332	0.63\\
341.18867066419	0.63\\
341.18867066419	0.64\\
341.188941915453	0.64\\
341.188941915453	0.65\\
341.233649784235	0.65\\
341.233649784235	0.66\\
341.319377633024	0.66\\
341.319377633024	0.67\\
341.367447208324	0.67\\
341.367447208324	0.68\\
341.461565805045	0.68\\
341.461565805045	0.69\\
341.694166635998	0.69\\
341.694166635998	0.7\\
341.979796768589	0.7\\
341.979796768589	0.71\\
342.091711610092	0.71\\
342.091711610092	0.72\\
342.399867866361	0.72\\
342.399867866361	0.73\\
342.705100766213	0.73\\
342.705100766213	0.74\\
342.711676025518	0.74\\
342.711676025518	0.75\\
342.808418737199	0.75\\
342.808418737199	0.76\\
342.868086561024	0.76\\
342.868086561024	0.77\\
343.204616950557	0.77\\
343.204616950557	0.78\\
343.380027715929	0.78\\
343.380027715929	0.79\\
343.412661092217	0.79\\
343.412661092217	0.8\\
343.870842816045	0.8\\
343.870842816045	0.81\\
343.996204986542	0.81\\
343.996204986542	0.82\\
344.223402900337	0.82\\
344.223402900337	0.83\\
344.313273778312	0.83\\
344.313273778312	0.84\\
344.846016619801	0.84\\
344.846016619801	0.85\\
344.934940167562	0.85\\
344.934940167562	0.86\\
344.943674489566	0.86\\
344.943674489566	0.87\\
345.042370478847	0.87\\
345.042370478847	0.88\\
345.326235894731	0.88\\
345.326235894731	0.89\\
346.824528353115	0.89\\
346.824528353115	0.9\\
347.191305676347	0.9\\
347.191305676347	0.91\\
347.844650361876	0.91\\
347.844650361876	0.92\\
348.548848045138	0.92\\
348.548848045138	0.93\\
348.557607438663	0.93\\
348.557607438663	0.94\\
348.81369061923	0.94\\
348.81369061923	0.95\\
350.560749637555	0.95\\
350.560749637555	0.96\\
351.844039925443	0.96\\
351.844039925443	0.97\\
351.950727891915	0.97\\
351.950727891915	0.98\\
354.506732955492	0.98\\
354.506732955492	0.99\\
358.397117441159	0.99\\
358.397117441159	1\\
inf	1\\
};
\addlegendentry{$\theta\!=\!0.5\sigma (\mathcal{L}\!=\!0.43$~y$)$}
\addplot [color=color3]
  table[row sep=crcr]{%
-inf	0\\
325.632828482502	0\\
325.632828482502	0.01\\
328.15810563317	0.01\\
328.15810563317	0.02\\
328.297494535968	0.02\\
328.297494535968	0.03\\
328.452413921744	0.03\\
328.452413921744	0.04\\
329.665939922045	0.04\\
329.665939922045	0.05\\
329.998764562832	0.05\\
329.998764562832	0.06\\
330.105444034886	0.06\\
330.105444034886	0.07\\
330.212005642252	0.07\\
330.212005642252	0.08\\
330.343845808845	0.08\\
330.343845808845	0.09\\
330.68602392043	0.09\\
330.68602392043	0.1\\
330.833497763105	0.1\\
330.833497763105	0.11\\
331.097613006915	0.11\\
331.097613006915	0.12\\
331.370343759543	0.12\\
331.370343759543	0.13\\
333.058641091628	0.13\\
333.058641091628	0.14\\
333.620882735272	0.14\\
333.620882735272	0.15\\
333.817281840297	0.15\\
333.817281840297	0.16\\
333.930140619954	0.16\\
333.930140619954	0.17\\
333.962408768643	0.17\\
333.962408768643	0.18\\
334.02471576927	0.18\\
334.02471576927	0.19\\
334.08105441433	0.19\\
334.08105441433	0.2\\
334.454792804531	0.2\\
334.454792804531	0.21\\
334.756296567694	0.21\\
334.756296567694	0.22\\
335.184460225381	0.22\\
335.184460225381	0.23\\
335.416804434728	0.23\\
335.416804434728	0.24\\
335.475481138663	0.24\\
335.475481138663	0.25\\
336.00933936885	0.25\\
336.00933936885	0.26\\
336.156569029895	0.26\\
336.156569029895	0.27\\
336.333129969804	0.27\\
336.333129969804	0.28\\
336.502455323494	0.28\\
336.502455323494	0.29\\
336.741567988815	0.29\\
336.741567988815	0.3\\
337.545702942587	0.3\\
337.545702942587	0.31\\
337.667546035878	0.31\\
337.667546035878	0.32\\
337.777917312333	0.32\\
337.777917312333	0.33\\
337.790476513706	0.33\\
337.790476513706	0.34\\
337.799634134223	0.34\\
337.799634134223	0.35\\
338.139071426041	0.35\\
338.139071426041	0.36\\
339.139154837478	0.36\\
339.139154837478	0.37\\
339.158085803626	0.37\\
339.158085803626	0.38\\
339.819232279984	0.38\\
339.819232279984	0.39\\
339.931912853804	0.39\\
339.931912853804	0.4\\
340.068316518022	0.4\\
340.068316518022	0.41\\
340.137449045842	0.41\\
340.137449045842	0.42\\
340.191121783518	0.42\\
340.191121783518	0.43\\
340.226097166389	0.43\\
340.226097166389	0.44\\
340.377983404762	0.44\\
340.377983404762	0.45\\
340.386016626953	0.45\\
340.386016626953	0.46\\
340.607169560588	0.46\\
340.607169560588	0.47\\
340.700899552437	0.47\\
340.700899552437	0.48\\
340.768629772058	0.48\\
340.768629772058	0.49\\
340.916341178536	0.49\\
340.916341178536	0.5\\
340.957854808634	0.5\\
340.957854808634	0.51\\
341.074471198613	0.51\\
341.074471198613	0.52\\
341.240662127564	0.52\\
341.240662127564	0.53\\
341.321285208329	0.53\\
341.321285208329	0.54\\
341.572949064933	0.54\\
341.572949064933	0.55\\
341.713009176396	0.55\\
341.713009176396	0.56\\
341.830407524153	0.56\\
341.830407524153	0.57\\
342.564643001258	0.57\\
342.564643001258	0.58\\
342.726056548258	0.58\\
342.726056548258	0.59\\
342.765947809957	0.59\\
342.765947809957	0.6\\
342.823515692634	0.6\\
342.823515692634	0.61\\
343.028111509418	0.61\\
343.028111509418	0.62\\
343.030447501354	0.62\\
343.030447501354	0.63\\
343.233097154139	0.63\\
343.233097154139	0.64\\
343.387204699049	0.64\\
343.387204699049	0.65\\
343.40706482008	0.65\\
343.40706482008	0.66\\
343.458608907825	0.66\\
343.458608907825	0.67\\
343.468442438281	0.67\\
343.468442438281	0.68\\
343.635928072464	0.68\\
343.635928072464	0.69\\
343.678885632945	0.69\\
343.678885632945	0.7\\
343.886865007939	0.7\\
343.886865007939	0.71\\
344.88217896368	0.71\\
344.88217896368	0.72\\
344.917182175458	0.72\\
344.917182175458	0.73\\
344.921363710459	0.73\\
344.921363710459	0.74\\
345.04344467147	0.74\\
345.04344467147	0.75\\
345.747873622417	0.75\\
345.747873622417	0.76\\
346.326857386437	0.76\\
346.326857386437	0.77\\
346.714146184739	0.77\\
346.714146184739	0.78\\
347.015734251379	0.78\\
347.015734251379	0.79\\
347.402872191229	0.79\\
347.402872191229	0.8\\
347.568519667226	0.8\\
347.568519667226	0.81\\
347.588957531875	0.81\\
347.588957531875	0.82\\
348.161072719187	0.82\\
348.161072719187	0.83\\
348.497000867896	0.83\\
348.497000867896	0.84\\
348.712480870704	0.84\\
348.712480870704	0.85\\
348.901173972104	0.85\\
348.901173972104	0.86\\
349.165937301663	0.86\\
349.165937301663	0.87\\
349.345184274409	0.87\\
349.345184274409	0.88\\
349.747813812286	0.88\\
349.747813812286	0.89\\
350.347200071328	0.89\\
350.347200071328	0.9\\
350.95863233794	0.9\\
350.95863233794	0.91\\
351.591054457437	0.91\\
351.591054457437	0.92\\
352.582767476643	0.92\\
352.582767476643	0.93\\
353.53714135085	0.93\\
353.53714135085	0.94\\
355.197478591275	0.94\\
355.197478591275	0.95\\
355.230227507839	0.95\\
355.230227507839	0.96\\
355.347155911348	0.96\\
355.347155911348	0.97\\
356.989197211972	0.97\\
356.989197211972	0.98\\
357.15442204145	0.98\\
357.15442204145	0.99\\
357.166098981931	0.99\\
357.166098981931	1\\
inf	1\\
};
\addlegendentry{$\theta\!=\!\sigma (\mathcal{L}\!=\!0.47$~y$)$}
\addplot [color=color4]
  table[row sep=crcr]{%
-inf	0\\
324.344165377861	0\\
324.344165377861	0.01\\
334.219117895806	0.01\\
334.219117895806	0.02\\
336.62324464314	0.02\\
336.62324464314	0.03\\
336.753985997661	0.03\\
336.753985997661	0.04\\
337.081327323354	0.04\\
337.081327323354	0.05\\
337.138761733381	0.05\\
337.138761733381	0.06\\
339.151692010164	0.06\\
339.151692010164	0.07\\
339.364135695431	0.07\\
339.364135695431	0.08\\
339.59746832268	0.08\\
339.59746832268	0.09\\
340.0869080685	0.09\\
340.0869080685	0.1\\
340.258038061356	0.1\\
340.258038061356	0.11\\
340.363626455247	0.11\\
340.363626455247	0.12\\
340.366268265202	0.12\\
340.366268265202	0.13\\
341.794525340838	0.13\\
341.794525340838	0.14\\
342.261639296561	0.14\\
342.261639296561	0.15\\
342.603212448703	0.15\\
342.603212448703	0.16\\
342.72666030886	0.16\\
342.72666030886	0.17\\
342.779343914364	0.17\\
342.779343914364	0.18\\
342.862720922732	0.18\\
342.862720922732	0.19\\
343.04190464757	0.19\\
343.04190464757	0.2\\
343.46944947187	0.2\\
343.46944947187	0.21\\
343.601806434676	0.21\\
343.601806434676	0.22\\
343.689406003522	0.22\\
343.689406003522	0.23\\
343.802940412269	0.23\\
343.802940412269	0.24\\
344.026599278441	0.24\\
344.026599278441	0.25\\
344.234629433931	0.25\\
344.234629433931	0.26\\
344.680767094439	0.26\\
344.680767094439	0.27\\
344.844598987864	0.27\\
344.844598987864	0.28\\
344.989670446428	0.28\\
344.989670446428	0.29\\
345.170205126577	0.29\\
345.170205126577	0.3\\
345.233171597366	0.3\\
345.233171597366	0.31\\
345.323864462323	0.31\\
345.323864462323	0.32\\
345.397451630929	0.32\\
345.397451630929	0.33\\
345.433613272124	0.33\\
345.433613272124	0.34\\
345.563181653401	0.34\\
345.563181653401	0.35\\
345.772607100541	0.35\\
345.772607100541	0.36\\
345.832231899814	0.36\\
345.832231899814	0.37\\
346.07317301258	0.37\\
346.07317301258	0.38\\
346.184191802842	0.38\\
346.184191802842	0.39\\
346.467600936837	0.39\\
346.467600936837	0.4\\
346.594537221646	0.4\\
346.594537221646	0.41\\
346.694319281752	0.41\\
346.694319281752	0.42\\
346.996241118223	0.42\\
346.996241118223	0.43\\
347.20043379305	0.43\\
347.20043379305	0.44\\
347.335092551169	0.44\\
347.335092551169	0.45\\
347.348200446347	0.45\\
347.348200446347	0.46\\
347.684578734011	0.46\\
347.684578734011	0.47\\
347.761707978507	0.47\\
347.761707978507	0.48\\
347.810380578786	0.48\\
347.810380578786	0.49\\
347.872616899363	0.49\\
347.872616899363	0.5\\
348.056716268094	0.5\\
348.056716268094	0.51\\
348.151356128185	0.51\\
348.151356128185	0.52\\
348.170579597115	0.52\\
348.170579597115	0.53\\
348.431148252449	0.53\\
348.431148252449	0.54\\
348.883452984281	0.54\\
348.883452984281	0.55\\
348.890745605745	0.55\\
348.890745605745	0.56\\
348.921945229635	0.56\\
348.921945229635	0.57\\
349.072272170353	0.57\\
349.072272170353	0.58\\
349.087628346281	0.58\\
349.087628346281	0.59\\
349.440572591327	0.59\\
349.440572591327	0.6\\
349.772706217495	0.6\\
349.772706217495	0.61\\
350.206796909546	0.61\\
350.206796909546	0.62\\
350.225813589744	0.62\\
350.225813589744	0.63\\
350.513420980257	0.63\\
350.513420980257	0.64\\
351.288275124799	0.64\\
351.288275124799	0.65\\
352.18282383122	0.65\\
352.18282383122	0.66\\
352.722708378238	0.66\\
352.722708378238	0.67\\
352.992905014623	0.67\\
352.992905014623	0.68\\
353.016116127442	0.68\\
353.016116127442	0.69\\
353.382579441618	0.69\\
353.382579441618	0.7\\
353.956778745987	0.7\\
353.956778745987	0.71\\
354.182689744788	0.71\\
354.182689744788	0.72\\
354.242501855251	0.72\\
354.242501855251	0.73\\
354.269354959829	0.73\\
354.269354959829	0.74\\
354.461443255813	0.74\\
354.461443255813	0.75\\
354.575507182516	0.75\\
354.575507182516	0.76\\
354.835371024933	0.76\\
354.835371024933	0.77\\
354.874850931078	0.77\\
354.874850931078	0.78\\
354.899462872879	0.78\\
354.899462872879	0.79\\
354.912600202384	0.79\\
354.912600202384	0.8\\
355.139831270958	0.8\\
355.139831270958	0.81\\
355.210775861744	0.81\\
355.210775861744	0.82\\
355.715408292607	0.82\\
355.715408292607	0.83\\
356.156444205912	0.83\\
356.156444205912	0.84\\
356.396754806503	0.84\\
356.396754806503	0.85\\
356.973009793274	0.85\\
356.973009793274	0.86\\
357.268307083345	0.86\\
357.268307083345	0.87\\
357.898493134538	0.87\\
357.898493134538	0.88\\
358.709532830356	0.88\\
358.709532830356	0.89\\
358.902747161912	0.89\\
358.902747161912	0.9\\
359.268208074606	0.9\\
359.268208074606	0.91\\
359.966500659775	0.91\\
359.966500659775	0.92\\
360.080508769356	0.92\\
360.080508769356	0.93\\
361.133871342578	0.93\\
361.133871342578	0.94\\
361.316125246543	0.94\\
361.316125246543	0.95\\
361.404043666335	0.95\\
361.404043666335	0.96\\
362.008925469334	0.96\\
362.008925469334	0.97\\
363.140620273162	0.97\\
363.140620273162	0.98\\
364.669278353005	0.98\\
364.669278353005	0.99\\
367.037327222302	0.99\\
367.037327222302	1\\
inf	1\\
};
\addlegendentry{$\theta\!=\!1.5\sigma (\mathcal{L}\!=\!0.52$~y$)$}
\addplot [color=color5]
  table[row sep=crcr]{%
-inf	0\\
344.984766738152	0\\
344.984766738152	0.01\\
345.001875835974	0.01\\
345.001875835974	0.02\\
346.173314535529	0.02\\
346.173314535529	0.03\\
347.817217715135	0.03\\
347.817217715135	0.04\\
347.89934545492	0.04\\
347.89934545492	0.05\\
348.573924964686	0.05\\
348.573924964686	0.06\\
350.681196682402	0.06\\
350.681196682402	0.07\\
352.091845020192	0.07\\
352.091845020192	0.08\\
352.165283087636	0.08\\
352.165283087636	0.09\\
352.95591946424	0.09\\
352.95591946424	0.1\\
353.385378843882	0.1\\
353.385378843882	0.11\\
354.505821780458	0.11\\
354.505821780458	0.12\\
354.623912839374	0.12\\
354.623912839374	0.13\\
354.791219964438	0.13\\
354.791219964438	0.14\\
355.854482440565	0.14\\
355.854482440565	0.15\\
356.160582756359	0.15\\
356.160582756359	0.16\\
356.380759948709	0.16\\
356.380759948709	0.17\\
356.659005365968	0.17\\
356.659005365968	0.18\\
356.943178544546	0.18\\
356.943178544546	0.19\\
357.123124888435	0.19\\
357.123124888435	0.2\\
358.376025780113	0.2\\
358.376025780113	0.21\\
358.427211096223	0.21\\
358.427211096223	0.22\\
358.530429135698	0.22\\
358.530429135698	0.23\\
358.565585912413	0.23\\
358.565585912413	0.24\\
358.651558565446	0.24\\
358.651558565446	0.25\\
358.656609492422	0.25\\
358.656609492422	0.26\\
359.429835154764	0.26\\
359.429835154764	0.27\\
359.600986384014	0.27\\
359.600986384014	0.28\\
359.648651717771	0.28\\
359.648651717771	0.29\\
359.931331664599	0.29\\
359.931331664599	0.3\\
360.007333535736	0.3\\
360.007333535736	0.31\\
360.340598251968	0.31\\
360.340598251968	0.32\\
360.368599377787	0.32\\
360.368599377787	0.33\\
360.371679320346	0.33\\
360.371679320346	0.34\\
360.535242259229	0.34\\
360.535242259229	0.35\\
360.695250831113	0.35\\
360.695250831113	0.36\\
361.782319178076	0.36\\
361.782319178076	0.37\\
361.966165799997	0.37\\
361.966165799997	0.38\\
361.974285453592	0.38\\
361.974285453592	0.39\\
362.019378249726	0.39\\
362.019378249726	0.4\\
362.190330497595	0.4\\
362.190330497595	0.41\\
362.755618994934	0.41\\
362.755618994934	0.42\\
362.777292446887	0.42\\
362.777292446887	0.43\\
363.164941212147	0.43\\
363.164941212147	0.44\\
363.227972571352	0.44\\
363.227972571352	0.45\\
363.343573363865	0.45\\
363.343573363865	0.46\\
363.519576281571	0.46\\
363.519576281571	0.47\\
363.542167214149	0.47\\
363.542167214149	0.48\\
363.732107981979	0.48\\
363.732107981979	0.49\\
363.882886855041	0.49\\
363.882886855041	0.5\\
363.989309372858	0.5\\
363.989309372858	0.51\\
364.049972561907	0.51\\
364.049972561907	0.52\\
364.306944528852	0.52\\
364.306944528852	0.53\\
364.93227312658	0.53\\
364.93227312658	0.54\\
365.061912788389	0.54\\
365.061912788389	0.55\\
365.232857558736	0.55\\
365.232857558736	0.56\\
365.29464904921	0.56\\
365.29464904921	0.57\\
365.455718468167	0.57\\
365.455718468167	0.58\\
365.825545633132	0.58\\
365.825545633132	0.59\\
365.954246258234	0.59\\
365.954246258234	0.6\\
366.797670780277	0.6\\
366.797670780277	0.61\\
366.99280436604	0.61\\
366.99280436604	0.62\\
367.402315283943	0.62\\
367.402315283943	0.63\\
367.603844847025	0.63\\
367.603844847025	0.64\\
367.664960984574	0.64\\
367.664960984574	0.65\\
367.773692013297	0.65\\
367.773692013297	0.66\\
367.862263839175	0.66\\
367.862263839175	0.67\\
367.864003423082	0.67\\
367.864003423082	0.68\\
368.087823162748	0.68\\
368.087823162748	0.69\\
368.321466114655	0.69\\
368.321466114655	0.7\\
368.629481770307	0.7\\
368.629481770307	0.71\\
369.14942339101	0.71\\
369.14942339101	0.72\\
369.685019197078	0.72\\
369.685019197078	0.73\\
370.020123936434	0.73\\
370.020123936434	0.74\\
370.111018061444	0.74\\
370.111018061444	0.75\\
370.26950987585	0.75\\
370.26950987585	0.76\\
370.30323119737	0.76\\
370.30323119737	0.77\\
370.579235319037	0.77\\
370.579235319037	0.78\\
370.619516941813	0.78\\
370.619516941813	0.79\\
370.630206201342	0.79\\
370.630206201342	0.8\\
370.982855209143	0.8\\
370.982855209143	0.81\\
371.514872046024	0.81\\
371.514872046024	0.82\\
371.644968491641	0.82\\
371.644968491641	0.83\\
371.951467339516	0.83\\
371.951467339516	0.84\\
372.14494238003	0.84\\
372.14494238003	0.85\\
372.584034835577	0.85\\
372.584034835577	0.86\\
373.273939622629	0.86\\
373.273939622629	0.87\\
373.537720352989	0.87\\
373.537720352989	0.88\\
373.861680279143	0.88\\
373.861680279143	0.89\\
374.745878219978	0.89\\
374.745878219978	0.9\\
376.21372961251	0.9\\
376.21372961251	0.91\\
376.24437594839	0.91\\
376.24437594839	0.92\\
378.249309091446	0.92\\
378.249309091446	0.93\\
378.891097852198	0.93\\
378.891097852198	0.94\\
379.301383104399	0.94\\
379.301383104399	0.95\\
380.001879449557	0.95\\
380.001879449557	0.96\\
380.329797189063	0.96\\
380.329797189063	0.97\\
381.723573113227	0.97\\
381.723573113227	0.98\\
382.045510640314	0.98\\
382.045510640314	0.99\\
384.300012881026	0.99\\
384.300012881026	1\\
inf	1\\
};
\addlegendentry{$\theta\!=\!2\sigma (\mathcal{L}\!=\!0.57$~y$)$}
\addplot [color=color6]
  table[row sep=crcr]{%
-inf	0\\
358.13979330692	0\\
358.13979330692	0.01\\
368.638049281908	0.01\\
368.638049281908	0.02\\
369.54363720101	0.02\\
369.54363720101	0.03\\
370.208000920324	0.03\\
370.208000920324	0.04\\
370.389770423731	0.04\\
370.389770423731	0.05\\
370.509879841331	0.05\\
370.509879841331	0.06\\
372.300514376448	0.06\\
372.300514376448	0.07\\
372.638682120453	0.07\\
372.638682120453	0.08\\
372.704864355371	0.08\\
372.704864355371	0.09\\
373.514821186975	0.09\\
373.514821186975	0.1\\
373.99194867252	0.1\\
373.99194867252	0.11\\
374.329056224504	0.11\\
374.329056224504	0.12\\
374.348822465931	0.12\\
374.348822465931	0.13\\
374.477887134799	0.13\\
374.477887134799	0.14\\
374.770857932633	0.14\\
374.770857932633	0.15\\
375.052766629622	0.15\\
375.052766629622	0.16\\
375.105012740687	0.16\\
375.105012740687	0.17\\
375.379670634333	0.17\\
375.379670634333	0.18\\
376.235080872842	0.18\\
376.235080872842	0.19\\
376.316536969348	0.19\\
376.316536969348	0.2\\
376.525023438225	0.2\\
376.525023438225	0.21\\
377.504566561266	0.21\\
377.504566561266	0.22\\
377.956457611974	0.22\\
377.956457611974	0.23\\
378.117445726573	0.23\\
378.117445726573	0.24\\
378.446942333998	0.24\\
378.446942333998	0.25\\
378.63430689707	0.25\\
378.63430689707	0.26\\
378.975346186015	0.26\\
378.975346186015	0.27\\
379.060486894086	0.27\\
379.060486894086	0.28\\
379.230171016077	0.28\\
379.230171016077	0.29\\
379.525294566048	0.29\\
379.525294566048	0.3\\
379.976505432684	0.3\\
379.976505432684	0.31\\
380.049224945555	0.31\\
380.049224945555	0.32\\
380.053208250405	0.32\\
380.053208250405	0.33\\
380.077293665745	0.33\\
380.077293665745	0.34\\
380.131683531844	0.34\\
380.131683531844	0.35\\
380.206363227623	0.35\\
380.206363227623	0.36\\
380.348487639569	0.36\\
380.348487639569	0.37\\
380.375049290124	0.37\\
380.375049290124	0.38\\
380.430856184382	0.38\\
380.430856184382	0.39\\
380.543913217605	0.39\\
380.543913217605	0.4\\
381.370576451208	0.4\\
381.370576451208	0.41\\
381.416824406233	0.41\\
381.416824406233	0.42\\
381.431967620112	0.42\\
381.431967620112	0.43\\
381.528614755838	0.43\\
381.528614755838	0.44\\
381.563078259332	0.44\\
381.563078259332	0.45\\
381.625789172651	0.45\\
381.625789172651	0.46\\
381.701036570281	0.46\\
381.701036570281	0.47\\
381.707732671251	0.47\\
381.707732671251	0.48\\
382.650559778033	0.48\\
382.650559778033	0.49\\
383.212471005164	0.49\\
383.212471005164	0.5\\
383.324042041371	0.5\\
383.324042041371	0.51\\
384.380818266197	0.51\\
384.380818266197	0.52\\
384.684663260802	0.52\\
384.684663260802	0.53\\
384.809760444378	0.53\\
384.809760444378	0.54\\
384.861141048444	0.54\\
384.861141048444	0.55\\
384.955100105666	0.55\\
384.955100105666	0.56\\
385.211777194595	0.56\\
385.211777194595	0.57\\
385.265115054922	0.57\\
385.265115054922	0.58\\
385.616407808643	0.58\\
385.616407808643	0.59\\
385.790954052566	0.59\\
385.790954052566	0.6\\
386.284883379458	0.6\\
386.284883379458	0.61\\
386.286108320041	0.61\\
386.286108320041	0.62\\
386.442064703319	0.62\\
386.442064703319	0.63\\
386.861583822563	0.63\\
386.861583822563	0.64\\
386.952328737443	0.64\\
386.952328737443	0.65\\
387.079110887354	0.65\\
387.079110887354	0.66\\
387.181550878624	0.66\\
387.181550878624	0.67\\
387.403619818828	0.67\\
387.403619818828	0.68\\
387.611033738775	0.68\\
387.611033738775	0.69\\
387.676014014587	0.69\\
387.676014014587	0.7\\
387.737455644223	0.7\\
387.737455644223	0.71\\
387.810649807752	0.71\\
387.810649807752	0.72\\
387.839711704062	0.72\\
387.839711704062	0.73\\
387.945885482592	0.73\\
387.945885482592	0.74\\
388.530970089308	0.74\\
388.530970089308	0.75\\
389.103705904992	0.75\\
389.103705904992	0.76\\
389.340128295772	0.76\\
389.340128295772	0.77\\
389.350130940168	0.77\\
389.350130940168	0.78\\
389.665744215349	0.78\\
389.665744215349	0.79\\
389.984032284678	0.79\\
389.984032284678	0.8\\
390.367196692599	0.8\\
390.367196692599	0.81\\
391.288464425252	0.81\\
391.288464425252	0.82\\
391.332999294995	0.82\\
391.332999294995	0.83\\
391.6004379104	0.83\\
391.6004379104	0.84\\
392.435142996557	0.84\\
392.435142996557	0.85\\
392.856144995148	0.85\\
392.856144995148	0.86\\
393.615755579487	0.86\\
393.615755579487	0.87\\
393.7837083484	0.87\\
393.7837083484	0.88\\
394.234394344401	0.88\\
394.234394344401	0.89\\
394.806267630016	0.89\\
394.806267630016	0.9\\
395.322312721821	0.9\\
395.322312721821	0.91\\
395.514744524842	0.91\\
395.514744524842	0.92\\
395.834622970451	0.92\\
395.834622970451	0.93\\
396.611504133933	0.93\\
396.611504133933	0.94\\
396.949101233381	0.94\\
396.949101233381	0.95\\
397.103153921731	0.95\\
397.103153921731	0.96\\
398.273279324735	0.96\\
398.273279324735	0.97\\
398.788760787581	0.97\\
398.788760787581	0.98\\
399.989200828692	0.98\\
399.989200828692	0.99\\
404.338677882722	0.99\\
404.338677882722	1\\
inf	1\\
};
\addlegendentry{$\theta\!=\!2.5\sigma (\mathcal{L}\!=\!0.63$~y$)$}
\addplot [color=color7]
  table[row sep=crcr]{%
-inf	0\\
389.387969952013	0\\
389.387969952013	0.01\\
390.003143192368	0.01\\
390.003143192368	0.02\\
390.40911919747	0.02\\
390.40911919747	0.03\\
390.71814838255	0.03\\
390.71814838255	0.04\\
392.667486211906	0.04\\
392.667486211906	0.05\\
394.081282110048	0.05\\
394.081282110048	0.06\\
394.108585741299	0.06\\
394.108585741299	0.07\\
395.196581347686	0.07\\
395.196581347686	0.08\\
395.495899093251	0.08\\
395.495899093251	0.09\\
395.599688087148	0.09\\
395.599688087148	0.1\\
395.998378143702	0.1\\
395.998378143702	0.11\\
396.168733753499	0.11\\
396.168733753499	0.12\\
396.21524374645	0.12\\
396.21524374645	0.13\\
396.898466639792	0.13\\
396.898466639792	0.14\\
398.471107668077	0.14\\
398.471107668077	0.15\\
398.498074953477	0.15\\
398.498074953477	0.16\\
399.20402248692	0.16\\
399.20402248692	0.17\\
399.346318748548	0.17\\
399.346318748548	0.18\\
399.519067304385	0.18\\
399.519067304385	0.19\\
399.538384464129	0.19\\
399.538384464129	0.2\\
399.984142464222	0.2\\
399.984142464222	0.21\\
400.108560844949	0.21\\
400.108560844949	0.22\\
400.283430702708	0.22\\
400.283430702708	0.23\\
401.308367064775	0.23\\
401.308367064775	0.24\\
401.341792163479	0.24\\
401.341792163479	0.25\\
401.586412442449	0.25\\
401.586412442449	0.26\\
403.075017726262	0.26\\
403.075017726262	0.27\\
403.337313674122	0.27\\
403.337313674122	0.28\\
403.616131028556	0.28\\
403.616131028556	0.29\\
403.925768516794	0.29\\
403.925768516794	0.3\\
404.003546210988	0.3\\
404.003546210988	0.31\\
404.221895053548	0.31\\
404.221895053548	0.32\\
404.494066859875	0.32\\
404.494066859875	0.33\\
404.593610676653	0.33\\
404.593610676653	0.34\\
404.872523613601	0.34\\
404.872523613601	0.35\\
405.492419158476	0.35\\
405.492419158476	0.36\\
405.583919255849	0.36\\
405.583919255849	0.37\\
405.59585873712	0.37\\
405.59585873712	0.38\\
405.684024196949	0.38\\
405.684024196949	0.39\\
406.010660129134	0.39\\
406.010660129134	0.4\\
406.079596985643	0.4\\
406.079596985643	0.41\\
406.122067515284	0.41\\
406.122067515284	0.42\\
406.391698579922	0.42\\
406.391698579922	0.43\\
406.416046886022	0.43\\
406.416046886022	0.44\\
406.456075218941	0.44\\
406.456075218941	0.45\\
406.669725438347	0.45\\
406.669725438347	0.46\\
407.032480818023	0.46\\
407.032480818023	0.47\\
407.475534483841	0.47\\
407.475534483841	0.48\\
407.483503754965	0.48\\
407.483503754965	0.49\\
407.657174714338	0.49\\
407.657174714338	0.5\\
407.785587405208	0.5\\
407.785587405208	0.51\\
407.929996962971	0.51\\
407.929996962971	0.52\\
407.998166566079	0.52\\
407.998166566079	0.53\\
408.519567464816	0.53\\
408.519567464816	0.54\\
408.687148108794	0.54\\
408.687148108794	0.55\\
409.570220486912	0.55\\
409.570220486912	0.56\\
409.639614787381	0.56\\
409.639614787381	0.57\\
409.860030294579	0.57\\
409.860030294579	0.58\\
410.166727463454	0.58\\
410.166727463454	0.59\\
410.46029131138	0.59\\
410.46029131138	0.6\\
411.094091146158	0.6\\
411.094091146158	0.61\\
411.95921296845	0.61\\
411.95921296845	0.62\\
412.077715259853	0.62\\
412.077715259853	0.63\\
412.368820883709	0.63\\
412.368820883709	0.64\\
412.554045539433	0.64\\
412.554045539433	0.65\\
412.572022756944	0.65\\
412.572022756944	0.66\\
412.603740306395	0.66\\
412.603740306395	0.67\\
412.631666439357	0.67\\
412.631666439357	0.68\\
412.947255339044	0.68\\
412.947255339044	0.69\\
413.507349111928	0.69\\
413.507349111928	0.7\\
413.838037828648	0.7\\
413.838037828648	0.71\\
414.452806253255	0.71\\
414.452806253255	0.72\\
414.599228030771	0.72\\
414.599228030771	0.73\\
414.600902284718	0.73\\
414.600902284718	0.74\\
414.781483984449	0.74\\
414.781483984449	0.75\\
414.917833154832	0.75\\
414.917833154832	0.76\\
415.617710467897	0.76\\
415.617710467897	0.77\\
415.971394361804	0.77\\
415.971394361804	0.78\\
416.116641155599	0.78\\
416.116641155599	0.79\\
416.481294143471	0.79\\
416.481294143471	0.8\\
417.609989262616	0.8\\
417.609989262616	0.81\\
417.695700316898	0.81\\
417.695700316898	0.82\\
417.838391743811	0.82\\
417.838391743811	0.83\\
417.929781215212	0.83\\
417.929781215212	0.84\\
418.087954384408	0.84\\
418.087954384408	0.85\\
418.342854442772	0.85\\
418.342854442772	0.86\\
418.670841410123	0.86\\
418.670841410123	0.87\\
419.463178081139	0.87\\
419.463178081139	0.88\\
419.477266503387	0.88\\
419.477266503387	0.89\\
420.484168727041	0.89\\
420.484168727041	0.9\\
421.162480845049	0.9\\
421.162480845049	0.91\\
421.497086340596	0.91\\
421.497086340596	0.92\\
421.917762038967	0.92\\
421.917762038967	0.93\\
422.549060514731	0.93\\
422.549060514731	0.94\\
422.794310577092	0.94\\
422.794310577092	0.95\\
423.022601460111	0.95\\
423.022601460111	0.96\\
423.856805586996	0.96\\
423.856805586996	0.97\\
426.766333801544	0.97\\
426.766333801544	0.98\\
427.53728751505	0.98\\
427.53728751505	0.99\\
433.042046804687	0.99\\
433.042046804687	1\\
inf	1\\
};
\addlegendentry{$\theta\!=\!3\sigma (\mathcal{L}\!=\!0.69$~y$)$}
\end{axis}

\end{tikzpicture}%

%% file: Images/Scenario_1_selected_1.tex
%
%
\begin{tikzpicture}

\begin{axis}[%
width=\pwidth,
height=\pheight,
at={(0.758in,0.481in)},
scale only axis,
bar width=0.8,
xmin=-0.2,
xmax=51.2,
xlabel style={font=\color{white!15!black}},
xlabel={Sensor index},
ymin=0,
ymax=0.75,
ylabel style={font=\color{white!15!black}},
ylabel={Polling Frequency},
axis background/.style={fill=white},
legend style={legend cell align=left, at={(0.985,0.97)},anchor=north east, align=left, legend columns=2,font={\footnotesize}, draw=white!15!black}
]
\addplot[ybar stacked, fill=color7, draw=black, area legend] table[row sep=crcr] {%
1	0\\
2	0.28819\\
3	0.00444\\
4	0.33561\\
5	0.19959\\
6	0.2012\\
7	0\\
8	0.26923\\
9	0.11905\\
10	0.35062\\
11	0\\
12	0.22431\\
13	0.02518\\
14	0.26914\\
15	0.19221\\
16	0.15637\\
17	0\\
18	0.26514\\
19	0.14366\\
20	0.44028\\
21	0\\
22	0.22666\\
23	0.02476\\
24	0.35115\\
25	0.209\\
26	0.15264\\
27	0\\
28	0.2515\\
29	0.12612\\
30	0.33688\\
31	0\\
32	0.18942\\
33	4e-05\\
34	0.29343\\
35	0.19937\\
36	0.16323\\
37	0\\
38	0.16398\\
39	0.11043\\
40	0.38337\\
41	0\\
42	0.148\\
43	0.00254\\
44	0.30605\\
45	0.19085\\
46	0.08915\\
47	0\\
48	0.26505\\
49	0.13756\\
50	0.23718\\
};
\addplot[forget plot, color=white!15!black] table[row sep=crcr] {%
-0.2	0\\
51.2	0\\
};
\addlegendentry{Update}

\addplot[ybar stacked, fill=color3, draw=black, area legend] table[row sep=crcr] {%
1	0\\
2	0.03507\\
3	0.001\\
4	0.08981\\
5	0.04934\\
6	0.03339\\
7	0\\
8	0.07585\\
9	0.0301\\
10	0.07288\\
11	0\\
12	0.06685\\
13	0.00622\\
14	0.04372\\
15	0.04733\\
16	0.05238\\
17	1e-05\\
18	0.0613\\
19	0.03438\\
20	0.10671\\
21	0\\
22	0.04879\\
23	0.00594\\
24	0.09281\\
25	0.04918\\
26	0.01919\\
27	0\\
28	0.07236\\
29	0.03008\\
30	0.07055\\
31	0\\
32	0.06136\\
33	0\\
34	0.06591\\
35	0.04662\\
36	0.05191\\
37	0\\
38	0.03504\\
39	0.02811\\
40	0.09887\\
41	0\\
42	0.04135\\
43	0.00069\\
44	0.08218\\
45	0.04524\\
46	0.02622\\
47	0\\
48	0.07351\\
49	0.0316\\
50	0.07357\\
};
\addplot[forget plot, color=white!15!black] table[row sep=crcr] {%
-0.2	0\\
51.2	0\\
};
\addlegendentry{Silent}

\end{axis}

\end{tikzpicture}%

%% file: Images/Scenario_1_selected_2.tex
%
%

\begin{tikzpicture}

\begin{axis}[%
width=\pwidth,
height=\pheight,
at={(0.758in,0.481in)},
scale only axis,
bar width=0.8,
xmin=-0.2,
xmax=51.2,
xlabel style={font=\color{white!15!black}},
xlabel={Sensor index},
ymin=0,
ymax=0.75,
ylabel style={font=\color{white!15!black}},
ylabel={Polling Frequency},
axis background/.style={fill=white},
legend style={legend cell align=left, at={(0.985,0.97)},anchor=north east, align=left, legend columns=2,font={\footnotesize}, draw=white!15!black}
]
\addplot[ybar stacked, fill=color7, draw=black, area legend] table[row sep=crcr] {%
1	0\\
2	0.14644\\
3	0.00316\\
4	0.25528\\
5	0.12941\\
6	0.10699\\
7	0\\
8	0.21247\\
9	0.05961\\
10	0.29811\\
11	0\\
12	0.18725\\
13	0.00505\\
14	0.23741\\
15	0.12922\\
16	0.14429\\
17	1e-05\\
18	0.21484\\
19	0.0649\\
20	0.29866\\
21	0\\
22	0.15937\\
23	0.00684\\
24	0.26289\\
25	0.13243\\
26	0.06745\\
27	0\\
28	0.21743\\
29	0.06382\\
30	0.31056\\
31	0\\
32	0.17727\\
33	0.00059\\
34	0.26639\\
35	0.12947\\
36	0.1478\\
37	0\\
38	0.18721\\
39	0.05236\\
40	0.30669\\
41	0\\
42	0.16631\\
43	0.00121\\
44	0.25305\\
45	0.12669\\
46	0.09767\\
47	0\\
48	0.22601\\
49	0.05867\\
50	0.2746\\
};
\addplot[forget plot, color=white!15!black] table[row sep=crcr] {%
-0.2	0\\
51.2	0\\
};
\addlegendentry{Update}

\addplot[ybar stacked, fill=color3, draw=black, area legend] table[row sep=crcr] {%
1	0\\
2	0.07745\\
3	0.00288\\
4	0.18094\\
5	0.07618\\
6	0.07783\\
7	0\\
8	0.15235\\
9	0.03511\\
10	0.14175\\
11	0\\
12	0.1367\\
13	0.00358\\
14	0.11505\\
15	0.07368\\
16	0.11964\\
17	0\\
18	0.12686\\
19	0.03688\\
20	0.20128\\
21	0\\
22	0.10368\\
23	0.00459\\
24	0.18334\\
25	0.07215\\
26	0.04666\\
27	0\\
28	0.15313\\
29	0.03626\\
30	0.14119\\
31	0\\
32	0.13143\\
33	0.00065\\
34	0.1309\\
35	0.07359\\
36	0.11815\\
37	0\\
38	0.09626\\
39	0.03171\\
40	0.19927\\
41	0\\
42	0.10354\\
43	0.00094\\
44	0.16931\\
45	0.07243\\
46	0.07266\\
47	0\\
48	0.15121\\
49	0.03339\\
50	0.12952\\
};
\addplot[forget plot, color=white!15!black] table[row sep=crcr] {%
-0.2	0\\
51.2	0\\
};
\addlegendentry{Silent}

\end{axis}

\end{tikzpicture}%

%% file: Images/Scenario_2_selected_1.tex
%
%
%
\begin{tikzpicture}

\begin{axis}[%
width=\pwidth,
height=\pheight,
at={(0.758in,0.481in)},
scale only axis,
bar width=0.8,
xmin=-0.2,
xmax=51.2,
xlabel style={font=\color{white!15!black}},
xlabel={Sensor index},
ymin=0,
ymax=0.75,
ylabel style={font=\color{white!15!black}},
ylabel={Polling Frequency},
axis background/.style={fill=white},
legend style={legend cell align=left, at={(0.985,0.97)},anchor=north east, align=left, legend columns=2,font={\footnotesize}, draw=white!15!black}
]
\addplot[ybar stacked, fill=color7, draw=black, area legend] table[row sep=crcr] {%
1	0.00421\\
2	0.12224\\
3	0.14466\\
4	0.16907\\
5	0.24648\\
6	0.06571\\
7	0.0855\\
8	0.16046\\
9	0.22331\\
10	0.25444\\
11	0.03213\\
12	0.16187\\
13	0.18976\\
14	0.20838\\
15	0.22314\\
16	0.10837\\
17	0.12449\\
18	0.17681\\
19	0.22846\\
20	0.23822\\
21	0.06468\\
22	0.18963\\
23	0.15863\\
24	0.23202\\
25	0.25356\\
26	0.0469\\
27	0.06642\\
28	0.17846\\
29	0.22034\\
30	0.24275\\
31	0.12093\\
32	0.09198\\
33	0.17429\\
34	0.24881\\
35	0.23774\\
36	0.03956\\
37	0.1573\\
38	0.16702\\
39	0.19328\\
40	0.24434\\
41	0.05063\\
42	0.12219\\
43	0.14061\\
44	0.19839\\
45	0.23992\\
46	0.04613\\
47	0.09032\\
48	0.17223\\
49	0.1827\\
50	0.25778\\
};
\addplot[forget plot, color=white!15!black] table[row sep=crcr] {%
-0.2	0\\
51.2	0\\
};
\addlegendentry{Update}

\addplot[ybar stacked, fill=color3, draw=black, area legend] table[row sep=crcr] {%
1	0.00101\\
2	0.03135\\
3	0.03939\\
4	0.04551\\
5	0.0593\\
6	0.01629\\
7	0.02217\\
8	0.04165\\
9	0.05423\\
10	0.05923\\
11	0.0084\\
12	0.03952\\
13	0.04815\\
14	0.05186\\
15	0.05806\\
16	0.0316\\
17	0.03329\\
18	0.04449\\
19	0.05467\\
20	0.05909\\
21	0.01696\\
22	0.04791\\
23	0.03994\\
24	0.05791\\
25	0.06325\\
26	0.01102\\
27	0.01826\\
28	0.04552\\
29	0.05387\\
30	0.05863\\
31	0.03012\\
32	0.02479\\
33	0.04334\\
34	0.05858\\
35	0.05954\\
36	0.00999\\
37	0.03956\\
38	0.0397\\
39	0.04912\\
40	0.06095\\
41	0.01379\\
42	0.03146\\
43	0.03488\\
44	0.04872\\
45	0.05835\\
46	0.01212\\
47	0.02294\\
48	0.04377\\
49	0.04525\\
50	0.06325\\
};
\addplot[forget plot, color=white!15!black] table[row sep=crcr] {%
-0.2	0\\
51.2	0\\
};
\addlegendentry{Silent}

\end{axis}

\end{tikzpicture}%

%% file: Images/Scenario_2_selected_2.tex
%
%
\definecolor{mycolor1}{rgb}{0.00000,0.44700,0.74100}%
\definecolor{mycolor2}{rgb}{0.85000,0.32500,0.09800}%
\begin{tikzpicture}

\begin{axis}[%
width=\pwidth,
height=\pheight,
at={(0.758in,0.481in)},
scale only axis,
bar width=0.8,
xmin=-0.2,
xmax=51.2,
xlabel style={font=\color{white!15!black}},
xlabel={Sensor index},
ymin=0,
ymax=0.75,
ylabel style={font=\color{white!15!black}},
ylabel={Polling Frequency},
axis background/.style={fill=white},
legend style={legend cell align=left, at={(0.985,0.97)},anchor=north east, align=left, legend columns=2,font={\footnotesize}, draw=white!15!black}
]
\addplot[ybar stacked, fill=color7, draw=black, area legend] table[row sep=crcr] {%
1	0.00337\\
2	0.07422\\
3	0.10041\\
4	0.12315\\
5	0.1911\\
6	0.0308\\
7	0.05646\\
8	0.12187\\
9	0.16348\\
10	0.18512\\
11	0.02261\\
12	0.10386\\
13	0.15366\\
14	0.15227\\
15	0.16664\\
16	0.09557\\
17	0.08894\\
18	0.12372\\
19	0.1903\\
20	0.18744\\
21	0.04336\\
22	0.15438\\
23	0.10763\\
24	0.17366\\
25	0.19949\\
26	0.01611\\
27	0.05384\\
28	0.14945\\
29	0.17126\\
30	0.18927\\
31	0.11379\\
32	0.07164\\
33	0.13566\\
34	0.20005\\
35	0.17882\\
36	0.04556\\
37	0.13094\\
38	0.11302\\
39	0.15065\\
40	0.19725\\
41	0.04074\\
42	0.09684\\
43	0.10803\\
44	0.15423\\
45	0.19184\\
46	0.03012\\
47	0.0613\\
48	0.14108\\
49	0.13115\\
50	0.21223\\
};
\addplot[forget plot, color=white!15!black] table[row sep=crcr] {%
-0.2	0\\
51.2	0\\
};
\addlegendentry{Update}

\addplot[ybar stacked, fill=color3, draw=black, area legend] table[row sep=crcr] {%
1	0.00311\\
2	0.05162\\
3	0.07205\\
4	0.08798\\
5	0.11534\\
6	0.02435\\
7	0.04169\\
8	0.08178\\
9	0.10442\\
10	0.11666\\
11	0.01703\\
12	0.07145\\
13	0.09786\\
14	0.09563\\
15	0.11042\\
16	0.07151\\
17	0.06269\\
18	0.08198\\
19	0.11322\\
20	0.11199\\
21	0.03414\\
22	0.10221\\
23	0.07271\\
24	0.10602\\
25	0.1197\\
26	0.01218\\
27	0.03908\\
28	0.09942\\
29	0.10055\\
30	0.11183\\
31	0.07074\\
32	0.05288\\
33	0.08714\\
34	0.1179\\
35	0.10591\\
36	0.03362\\
37	0.08178\\
38	0.0731\\
39	0.09324\\
40	0.12106\\
41	0.03162\\
42	0.06242\\
43	0.06998\\
44	0.09677\\
45	0.11361\\
46	0.02413\\
47	0.04184\\
48	0.08701\\
49	0.08604\\
50	0.12021\\
};
\addplot[forget plot, color=white!15!black] table[row sep=crcr] {%
-0.2	0\\
51.2	0\\
};
\addlegendentry{Silent}

\end{axis}

\end{tikzpicture}%